\documentclass[useAMS,usenatbib]{mnras}

\usepackage{graphicx,epstopdf}	
\usepackage{multicol}
\usepackage{amsmath}   
\usepackage{lscape}
\usepackage{array}
\usepackage{lineno}
\usepackage{rotating} 
\usepackage{times}
\usepackage{subfig}
\usepackage{mathrsfs}

\newcommand{\kms}{km{\scriptsize /}s}

\newcommand{\mhi}{\mbox{$M_{\rm HI}$}}
\newcommand{\mhis}{\mbox{$M^*_{\rm HI}$}}
\newcommand{\msol}{\mbox{${\rm M}_\odot$}}
\newcommand{\thetas}{\mbox{$\theta^*$}}
\makeatletter
\newcommand*{\centerfloat}{%
  \parindent \z@
  \leftskip \z@ \@plus 1fil \@minus \textwidth
  \rightskip\leftskip
  \parfillskip \z@skip}
\makeatother

\title{The Ursa Major Association of Galaxies. VI: A relative dearth of gas-rich dwarf galaxies}
\author[Busekool et al.]{E. Busekool$^{1}$, M.A.W. Verheijen$^{1}$\thanks{E-mail: verheyen@astro.rug.nl},  J.M. van der Hulst$^{1}$, R.B. Tully$^{2}$,
\newauthor
N. Trentham$^{3, 4}$, and M.A. Zwaan$^{5}$ \\
$^{1}$Kapteyn Astronomical Institute, University of Groningen, P.O. Box 800, 9700 AV Groningen, The Netherlands \\
$^{2}$Institute for Astronomy (IFA), University of Hawaii, 2680 Woodlawn Drive, HI 96822, USA \\
$^{3}$Institute of Astronomy, Madingley Road, Cambridge CB3 0HA, UK \\
$^{4}$Deceased \\
$^{5}$European Southern Observatory, Karl-Schwarzschild-Strasse 2, D-85748, Garching, Germany} 
\begin{document}
\maketitle

\begin{abstract}
We determined the HI mass function of galaxies in the Ursa Major association of galaxies using a blind VLA-D array survey, consisting of 54 pointings in a cross pattern, covering the centre as well as the outskirts of the Ursa Major volume. The calculated HI mass function has best-fitting Schechter parameters $\theta^* = 0.19 \pm 0.11$ Mpc$^{-3}$, log $\mhis\ /M_{\odot} = 9.8 \pm 0.8$, and $\alpha = -0.92 \pm 0.16$. The high-mass end is determined by a complementary, targeted WSRT survey, the low-mass end is determined by the blind VLA survey. The slope is significantly shallower than the slopes of the HIPASS ($\alpha=-1.37\pm 0.03\pm 0.05$) and ALFALFA ($\alpha=-1.33\pm 0.02$) HI mass functions, which are measured over much larger volumes and cover a wider range of cosmic environments: There is a relative lack of low HI mass galaxies in the Ursa Major region. This difference in the slope strongly hints at an environmental dependence of the HI mass function slope.
\end{abstract}


\section{Introduction}

One of the big questions in astronomy is how galaxies form and evolve. The main theory of galaxy formation and evolution is based on the $\Lambda$CDM cosmological model of hierarchical galaxy formation. This model describes the formation of dark matter haloes from the distribution of primordial density fluctuations and predicts an evolving dark matter halo Mass Function (MF) with a dependence on cosmic environment \citep{jenkins2001}. Semi-Analytic Models (SAMs) (\citet{hirschmann2012}; \citet{fu2013}; \citet{merson2013}; \citet{lu2014}; \citet{somerville2015}) and hydrodynamical models (\citet{springel2010a}; \citet{springel2010b}; \citet{schaye2015}) describe how baryonic matter is accumulated in these dark matter haloes: gas flows into the dark matter haloes from the cosmic web, settles and forms stars: the creation of a galaxy. Low-mass galaxies merge subsequently and, in combination with gas accretion \citep{sancisi2008}, give rise to the build-up of larger galaxies, Milky Way size or bigger (e.g. \citet{davies1985}).

The slope of the predicted dark matter halo MF is much steeper than the slopes of the observed galaxy Luminosity Function (LF) \citep{blanton2001} and HI Mass Function (HIMF) (e.g. \citet{zwaan2005}; \citet{martin2010}). To cure this discrepancy, the SAMs and hydrodynamical models implemented different mechanisms using the observed LF and the HIMF as constraints. At the low-mass end, feedback is mainly driven by stellar feedback including supernova winds which produce an outflow of the gas from a galaxy (e.g. \citet{benson2003}; \citet{weinmann2012}). Other ideas about an alteration of the slope of the low-mass end of the LF and HIMF are: (i) low-mass haloes being inefficient in collecting baryons \citep{tully2002}; (ii) most of the baryons in low-mass haloes, which formed later in the lower density regions, may not yet have been converted into stars \citep{roychowdhury2012}; (iii) low-mass haloes are affected more by gas removal mechanisms, such as ram pressure stripping \citep{springob2005}. An important question is: which of these processes dominate and is there an environmental dependence?

In order to answer this question, one needs to determine the LF and the HIMF such that the effect of selection biases is minimized and systematic uncertainties are reduced. To avoid the uncertainties related to optical selection effects, blind HI spectral line surveys are a good way to construct an HIMF. The HIMF is usually parameterized with a Schechter function \citep{schechter1976} with three parameters; \thetas\ and \mhis\ are describing the knee while the slope is described by $\alpha$. The first significant blind survey was the Arecibo HI Strip Survey (AHISS) \citep{zwaan1997}, containing 65 detected galaxies. The HIMF of this sample has a slope of $\alpha \sim -1.2 \pm 0.1$. More recent blind HI surveys with single-dish telescopes are the HI Parkes All Sky Survey (HIPASS) \citep{zwaan2005} and the Arecibo Legacy Fast Arecibo L-band Feed Array (ALFALFA) survey \citep{martin2010}. 

These HI surveys have HI mass ranges of respectively $10^{7.0-10.6}$\msol\  \citep{zwaan2003} and $10^{6.2-10.8}$\msol\ \citep{haynes2011}, with the lower mass limit valid for a distance of 3 Mpc. All parameters of the HIMFs are listed in Table \ref{tab:HIMFs} and  illustrated in Figure 1. There is a significant difference in the \mhis\ of the HIPASS and ALFALFA surveys, the latter being higher than the former. This difference is not yet understood. The number density, \thetas, and the slopes of the HIMFs of HIPASS ($\alpha = -1.37$) and ALFALFA ($\alpha = -1.33$) are consistent with each other after renormalisation for the same cosmology. The number density of galaxies is increasing towards low-mass galaxies.  These slopes hint at hierarchical galaxy formation, as there are many more low-mass objects than high-mass objects \citep{navarro1997}.  

\begin{table*}
\begin{minipage}[c]{230mm}
\caption{The parameters of HI mass functions from various recent surveys.}\label{tab:HIMFs}
\begin{tabular}{lccccccc}

Survey & \# of & $\Theta^{*}$ & log$(M_{HI}^{*}/M_{\odot})$  & $\alpha$ & max. vel. & sky area & volume\\
 &galaxies & $10^{-3}$ Mpc$^{-3}$ dex$^{-1}$ & & & km/s & deg$^2$ & Mpc$^3$\\
\hline
\multicolumn{7}{l}{Large area surveys} \\
\hline
AHISS$^{(1)}$ & 66 & 5,6 & 9.81 & $-1.20 \pm 0.1$ & 7,400 & 65 & 6,469\\
HIPASS$^{(2)}$ & 4,315 & $5.8 \pm 0.8 \pm 0.6$ & $ 9.81 \pm 0.03 \pm 0.03$ & $-1.37 \pm 0.03 \pm 0.05$ & 12,700 & 21,341 & $1.0 \times$$10^8$\\
ALFALFA 40\%$^{(3)}$ & 10,119 & $5.7 \pm 0.3$ &$ 9.91 \pm 0.02$& $-1.33 \pm 0.02 $ & 18,000 & 2,607 & $3.6 \times$$10^7$\\
Springob$^{(4)}$ & 2,771 & $3.8$ &$ 9.94 $& $-1.24 \pm 0.1$ & 28,000 & all sky$^{(9)}$ & all sky$^{(9)}$\\
\hline
\multicolumn{7}{l}{Specific environment blind surveys} \\
\hline
6 groups$^{(5)}$ & 31 & \multicolumn{2}{c}{consistent with HIPASS} & -1.00 & 1,000 & 1 Mpc$^2$ & 13\\
8 groups$^{(6)}$ & 54 & \multicolumn{2}{c}{consistent with HIPASS} & -1.00 & $660$$-$$1,100^{(10)}$ & $\pm 1^{(10)}$ & $-$ \\
Canes Venatici$^{(7)}$ & 70 & 106$\pm$43 & $9.52 \pm 0.26$ &  $-1.17 +0.07/-0.08$ & 1, 330 & 86 & 62\\
Leo I$^{(8)}$ & 65 & $-$ & $-$ & $-1.41+ 0.2/- 0.1$ & 1,200 & 118 & 19\\
\hline
\multicolumn{8}{l}{(1) \citet{zwaan1997} (2) \citet{zwaan2005} (3) \citet{martin2010} (4) \citet{springob2005} (5) \citet{pisano2011} (6) \citet{freeland2009} } \\
\multicolumn{8}{l}{(7) \citet{kovac2007} (8) \citet{stierwalt2009} (9) optically selected (10) per group} \\
\end{tabular}
\end{minipage}
\end{table*}

\begin{figure}
\includegraphics[width=0.5\textwidth]{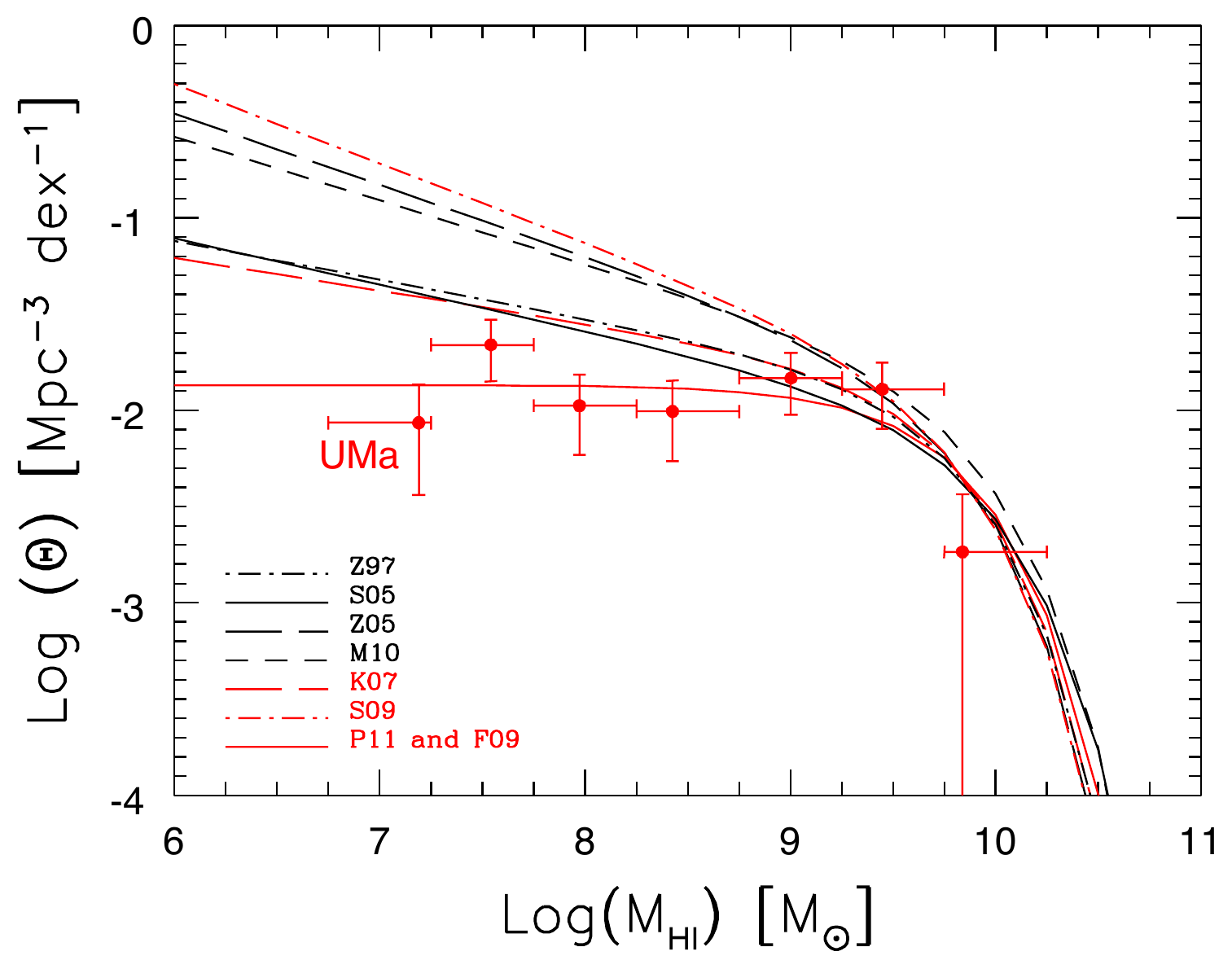}
\caption{Different HI mass functions. HIMFs of the large area surveys are black (dashed-pointed curve: \citet{zwaan1997}, solid curve: \citet{springob2005}, long-dashed curve: \citet{zwaan2005}, and short-dashed curve: \citet{martin2010}) and the HIMFs of specific environments are red (long-dashed curve: \citet{kovac2007}, dashed-pointed curve: \citet{stierwalt2009}, and solid curve: composite groups (\citet{pisano2011} and \citet{freeland2009})). All parameters are rescaled to H$_{0}$ = 74 km/s Mpc$^{-1}$ and have been renormalised to overlap around \mhis. The HIMF of UMa from this study (see Figure \ref{fig:HIMFUMa}) is plotted for comparison. The difference between the large, blind surveys and two composite groups is clearly visible.}\label{fig:diffHIMF}
\end{figure}

The effect of the environment on the HIMF can be studied by using large surveys and by comparing observations of specific environments. Large area surveys are used by \citet{springob2005} and \citet{zwaan2005}. \citet{springob2005} selected from the Arecibo General Catalogue (AGC) catalogue \citep{springob2005b} and calculated the local cosmic density of galaxies for each source by using data from the IRAS Point Source Catalogue of galaxies with a redshift (PSCz) \citep{saunders2000} and found that the slope of the HIMF is steeper in regions with a lower galaxy density. \citet{zwaan2005} constructed the HIMFs of different environments using the HIPASS catalogue HICAT and calculated the local cosmic density on the basis of the HIPASS catalogue (HICAT) itself by considering the distance from the first to the fifth nearest neighbour. They found that the slope of the HIMF is steeper in regions with high density, contrary to what \citet{springob2005} had found. \citet{springob2005} however consider their result non-conclusive, since within the errors, the HIMFs in different environments could have the same slope. Although the results appear contradictory, neither is conclusive. 

Another approach to investigate an environmental dependence, is to deliberately target and observe different cosmic environments and measure the HIMF in each environment separately (see Figure 1 and Table \ref{tab:HIMFs}). \citet{pisano2011} determined the slope of the HIMF for a combined sample of 31 galaxies in six galaxy groups and \citet{freeland2009} determined this for 54 galaxies in eight groups; both found that the slopes in galaxy groups are consistent with being flat. It should be noted, however, that stacking groups will make the volume corrections more complicated. In another survey, \citet{kovac2007} found that the slope of the HIMF of the nearby dwarf-dominated Canes Venatici groups is $-1.17 \pm 0.04$ with galaxies in the mass range of \mhi = $10^{6.5-10.5}$ M$_{\odot}$. \citet{stierwalt2009} used the ALFALFA survey to extract the HIMF of 65 galaxies in the Leo I group which consists of many low-surface brightness (LSB) gas-rich dwarfs and a few $L^{*}$ galaxies with a mass ranges of \mhi = $10^{6.5-10.0}$ M$_{\odot}$. The slope of this HIMF is $\alpha = -1.41 \pm 0.2$, which is very steep compared to other HIMFs, but with a large error. According to Table \ref{tab:HIMFs} CVn and Leo I are comparable surveys, however, their slopes do not match. An explanation for this could be the difference in environment or evolutionary stage \citep{stierwalt2009}. Besides this, the Leo I sample does not provide enough detections to fit all parameters of the Schechter function, hence only the slope is fitted. Changing \mhis and \thetas could have a huge effect on the fitted slope. From these results, it is clear that determining the environmental effect on the slope of the low-mass end of the HIMF is difficult, since the systematic errors in the calculated slopes are still very large and low-number statistics dominate the errors.

Most surveys determining the HIMF are volume and HI flux-limited, and the errors on their lowest HI masses are large, mainly due to distance uncertainties. These arise because low-HI mass galaxies are only detectable in the nearby universe where the Hubble flow cannot be used to determine their distances based on recession velocities. Consequently, the high-mass end of the HIMF may be well defined, although the difference in \mhis\ between HIPASS and ALFALFA suggests otherwise. The slope at the low-mass end is not so well defined as outlined in the previous paragraph. Another effect that has to be taken into account is that if the local volume dominates the sample, the galaxy population and density may become unrepresentative. To better constrain the slope of the low-mass end of the HIMF, blind and deep surveys of more distant regions with large numbers of galaxies may help. For such a survey the Ursa Major (UMa) region, dominated by gas-rich, late-type galaxies is an excellent target. The volume comprises a reasonably uniform environment within a coherent large scale structure. It is at a distance of 17.4 Mpc \citep{tully2012}, sufficiently nearby to detect dwarf galaxies in HI and, if the HIMF from ALFALFA or HIPASS is valid within this volume, we can expect hundreds of HI detections above a reasonable detection limit. Since all galaxies are roughly at the same distance, the relative distance uncertainties are small, thus the lowest HI masses can be calculated with relatively high accuracy which is ideal for fitting the low-mass end. The high-mass end can be constrained by pointed HI observations with the WSRT of all galaxies within the UMa region brighter than $M_{B} < -16.8$ \citep{verheijen2001b}. This study combines the merits of having a \textit{blind} survey with a \textit{well-defined} environment and adds a valuable, independent measurement of the slope of the HIMF. A further description of the UMa volume and the enclosed galaxy population is provided in Section \ref{sec:UMavol}.

In this paper we report on the slope of the low-mass end of the HIMF of the Ursa Major region. This paper is structured as follows. We describe the Ursa Major collection of galaxies in Section 2 and the details of the blind HI survey including the data reduction in Section 3. The observational results are presented in Section 4, the volume corrections and the calculations of the HIMF are described in Sections 5 and 6 respectively. Section 7 discusses a comparison with the ALFALFA results, followed by a discussion in Section 8. Throughout this work we assume H$_{0}$ = 74 km s$^{-1}$ Mpc$^{-1}$.

\section{The Ursa Major volume}\label{sec:UMavol}

\subsection{Sample description}

The Ursa Major (UMa) volume comprises a significant overdensity of galaxies located within the Virgo supercluster in the supergalactic plane. It spans a circular area on the sky of 15 degrees in diameter, centered on 11$^{h}$59$^{m}$28$^{s}$.3, 49$^{\circ}$05$^{'}$18$^{''}$ with a galactocentric velocity range of 700-1210 km/s \citep{tully1996}. \citet{tully1996} identified 79 members from the nearby galaxy catalogue (\citet{tully1988}; NBG). The members do not display any concentration toward a core. The sample consists mostly of spirals and late-type systems (83 per cent), contains a dozen lenticulars (15 per cent) and maybe two dwarf ellipticals (2 per cent). The velocity dispersion of this sample is 148 km s$^{-1}$ \citep{trentham2001}, and the crossing time is comparable to the Hubble time. \citet{wolfinger2013} found 51 HI detections of 96 optical sources in the HIJASS catalogue in the UMa region. \citet{karachentsev2013} has studied the region as well and concluded that the distribution of galaxies is patchy and elongated along the line of the supergalactic equator. The volume contains seven groups  and seems to be more an association of groups than a cluster. \citet{pak2014} have identified a total of 166 galaxies with $94$ per cent completeness by a NASA/IPAC Extragalactic Database (NED) and Sloan Digitized Sky Survey (SDSS) catalogue search. Their study confirms that the galaxy population of the UMa region is dominated by late-type galaxies and the existence of several subgroups is discussed. From the Rosat\footnote{http://heasarc.gsfc.nasa.gov/cgi-bin/W3Browse/w3browse.pl} image of this region we conclude that no diffuse X-ray radiation is detected, suggesting that no significant intergalactic medium is present against which ram-pressure stripping or strangulation may result in a depleted population of low-mass, gas-rich galaxies. Hence, we conclude that this region presents a well-defined environment to study the low-mass end of the HI mass function.

A large amount of photometric and HI synthesis imaging data already exists for galaxies in the UMa region. \citet{tully1996} present B, R and I-band surface photometry for all 79 members identified at that time, as well as near-infrared K' band surface photometry for a complete subsample of 62 galaxies, brighter than $M_{B} < -16.5$. For 56 galaxies in this subsample, cold atomic HI gas has been detected at 21cm with single dish and interferometric radio telescopes. Hence, in principle, the HI completeness of this optically complete subsample is 90 per cent. HI synthesis imaging data for 44 of these 56 galaxies are presented by \citet{verheijen2001b}. These were selected to be more inclined than 45$^{\circ}$ for the purpose of kinematic studies while 13 contain a warp, more than half is lopsided and 4 pairs are interacting with two companions confused in the single dish spectra. HI gas was detected in 10 of the 17 galaxies fainter than the optical completion limit ($M_{B} > -16.5$) while 8 of those have an HI mass more than an order of magnitude below the 'knee' mass of the ALFALFA HIMF. These 10 faint galaxies are inconsequential for our analysis and conclusions as the optically selected, targeted HI observations described here will only be used to constrain the high mass end of the HIMF in Ursa Major.

\citet{trentham2001} present results from an extensive wide-field R-band imaging survey aimed at measuring the faint-end slope of the luminosity function. This optical survey was done in conjunction with the blind HI survey presented in this paper. They found a flat slope ($\alpha \sim -1.00$) for the LF of the UMa region.

\subsection{The line-of-sight distribution of galaxies}

Understanding the distribution of galaxies within the UMa volume is important to be able to make appropriate volume corrections when calculating the HIMF. Different approaches to determine this distribution can be explored. Before we do so, the distance to the region, the depth, and the volume itself need to be determined.

The ensemble-average distance of the UMa galaxies is 17.4 Mpc as determined by \citet{tully2012} using the Tully-Fisher (TF) relation combined with distances based on the tip of the red giant branch (TRGB) and Cepheids. 

The most direct approach to determine the depth of the volume is to assume a quiet Hubble-flow and neglect peculiar motions. The galactocentric velocity window of the region (700-1210 km s$^{-1}$) would correspond to a depth range of 9.5-16.4 Mpc (H$_0$=74 km s$^{-1}$ Mpc$^{-1}$). However, its recession velocity is affected by the velocity field of the Local Supercluster which is dominated by the Virgo cluster as illustrated in Cosmic Flows 2 \footnote{http://irfu.cea.fr/cosmography} \citep{courtois2013}. The UMa volume is located at a physical distance of $8-13$ Mpc from the core of the Virgo cluster and, consequently, significant velocity crowding occurs in the direction of the UMa cluster. When the peculiar velocity of the region (-386 km s$^{-1}$ \citep{lavaux2010} \footnote{Extra galactic Distance Database (http://www.ifa.hawaii.edu/cosmicflows/)}) is taken into account, the depth range becomes 14.7-21.6 Mpc, which is quite different from the previous estimate assuming a quiet Hubble flow and more in line with the average distance of the galaxies in the region, based on Cepheids and the TRGB.

An estimate of the depth can also be obtained by taking a mean distance of 17.4 Mpc and assuming that the distribution of galaxies along the line-of-sight is similar to the somewhat elongated distribution on the sky along supergalactic longitude, this suggests a depth range of 15.1-19.7 Mpc. 

The last method we have explored to determine the depth is using the scatter in the TF relation. \citet{verheijen2001a} investigated the scatter in the TF relation of the UMa sample and found that the tightest correlation follows when using K'-band magnitudes and the amplitudes of the outer flat parts of the HI rotation curves. The total observed scatter is 0.26 mag and is a combination of measurement errors, intrinsic scatter, and the scatter due to the depth of the volume ($\sigma_{tot} = \sqrt{\sigma^2_{meas}+\sigma^2_{intrinsic}+\sigma^2_{depth}}$). If we take the measurement uncertainties and the slope of the TF relation into account, and assume $\sigma_{intrinsic} = 0$ mag, then all of the remaining excess scatter of 0.22 mag can be attributed to the depth of the UMa volume. Given the above-mentioned distance of 17.4 Mpc and assuming a Gaussian distribution of the galaxies along the line-of-sight, this translates to a $\pm$2.5-sigma depth range of 13.6-22.6 Mpc. Any unknown, non-zero intrinsic scatter would reduce this depth.

From the three methods described above, method two and three are not independent since both are using the average distance. However, method one is independent from the other two, and all methods give three depths, hence three volumes, of the UMa region which are in agreement. Within these three volumes, galaxies can be distributed in different ways, and this can influence the calculated HIMF. In this paper we will assume that the galaxies are distributed uniformly throughout the volume, motivated by a flat velocity distribution (\citet{tully1996}; \citet{pak2014}, see also Figure \ref{fig:W20vsVhel}). 

We adopt the depth range of 14.7-21.6 Mpc based on Cosmic Flows 2, with an uniform distribution of galaxies within the volume to calculate volume corrections, and an average ensemble distance of 17.4 Mpc to calculate the \mhi\ of the detections throughout this paper unless stated otherwise. This depth range and the 15 degree wide circular area on the sky define a total conical volume of the region of 121 Mpc$^3$.

\subsection{Expectations}

\begin{figure}
\includegraphics[width=0.5\textwidth]{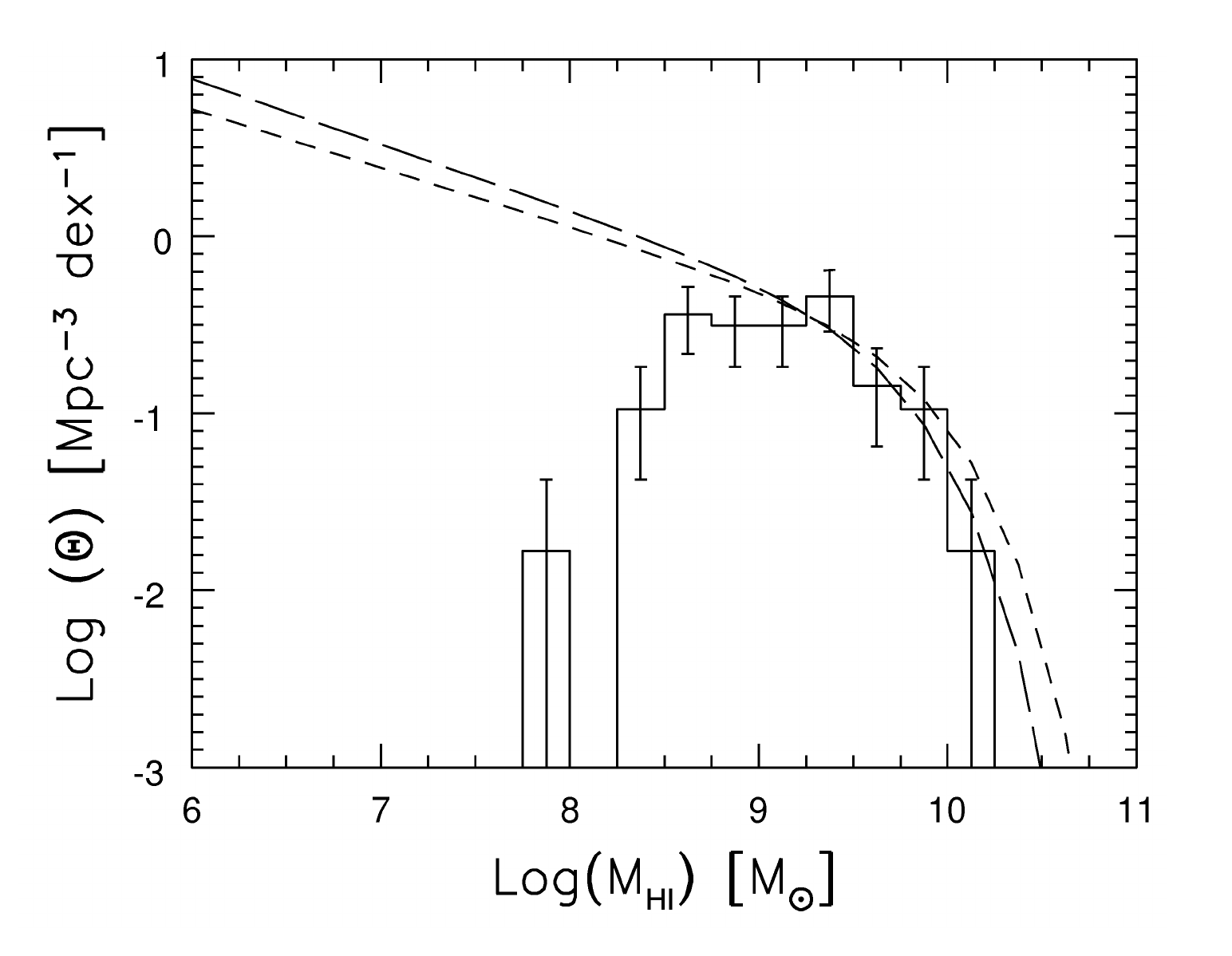}
\caption{The solid histogram shows the observed distribution of HI masses in the Ursa Major cluster based on an optically selected, complete sample. No volume corrections are applied. The short-dashed line is the renormalized HIMF of the ALFALFA survey with a slope of -1.33. The long-dashed curve is the renormalized HIMF of the HIPASS survey with a slope of -1.37. Both \thetas\ are set to match the WSRT sample.}\label{fig:histHA}
\end{figure}

Figure \ref{fig:histHA} shows a histogram of the observed WSRT HI mass function (HIMF) of the UMa sample, without any volume or completeness corrections, based on the sample of optically selected cluster members as presented in \citet{tully1996}.

The \thetas\ of the WSRT sample is normalized by the total volume argued above. The HIPASS and ALFALFA HIMF are plotted as well, normalized to the complete HI WSRT sample for comparison. The complete sample has a completion limit of $M_{B} < -16.5$, which corresponds to \mhi $\sim 10^{8.5}$ according to the \mhi/$L_{B}$ ratio. From this we assume that the WSRT HI sample is complete above an HI mass-limit of  \mhi\ $> 10^9$ \msol\, which is low enough to accurately fit \thetas. For HIPASS, the normalisation results in \thetas\ =  0.11 Mpc$^{-3}$ dex$^{-1}$ and for ALFALFA \thetas\ = 0.13 Mpc$^{-3}$ dex$^{-1}$. The short-dashed line is the HIMF from the ALFALFA survey with a slope of -1.33 and the long-dashed curve is the HIMF from the HIPASS survey with a slope of -1.37. From the renormalization of \thetas\ we conclude that the UMa cluster is overdense by a factor of about 18 compared to the average cosmic density as determined by HIPASS and ALFALFA. An accurate measurement of the overdensity is essential for obtaining a sufficient detection rate at the low-mass end.

Given the renormalized HIMFs of HIPASS and ALFALFA in Figure \ref{fig:histHA}, one can ask how large the expected population of dwarf galaxies in the UMa volume actually would be, assuming that galaxies with an HI mass as small as 10$^7$ M$_\odot$ could be detected throughout the entire UMa volume and that these galaxies are distibuted uniformly in space. Assuming an HIMF with a slope of $-1.35$ and log$(M_{HI}^{*}/M_{\odot})$ of 9.88 (the average of the HIPASS and ALFALFA values), with \thetas\ normalised to the WSRT sample as illustrated in Figure\ref{fig:histHA}, and integrating that HIMF upward from 10$^7$ M$_\odot$, would yield a population of 688 galaxies in the total volume. Even if only a fraction of the volume would be observed, a sufficient number of HI detections will be obtained to determine the slope of the HIMF in UMa.

\begin{figure*}
\includegraphics{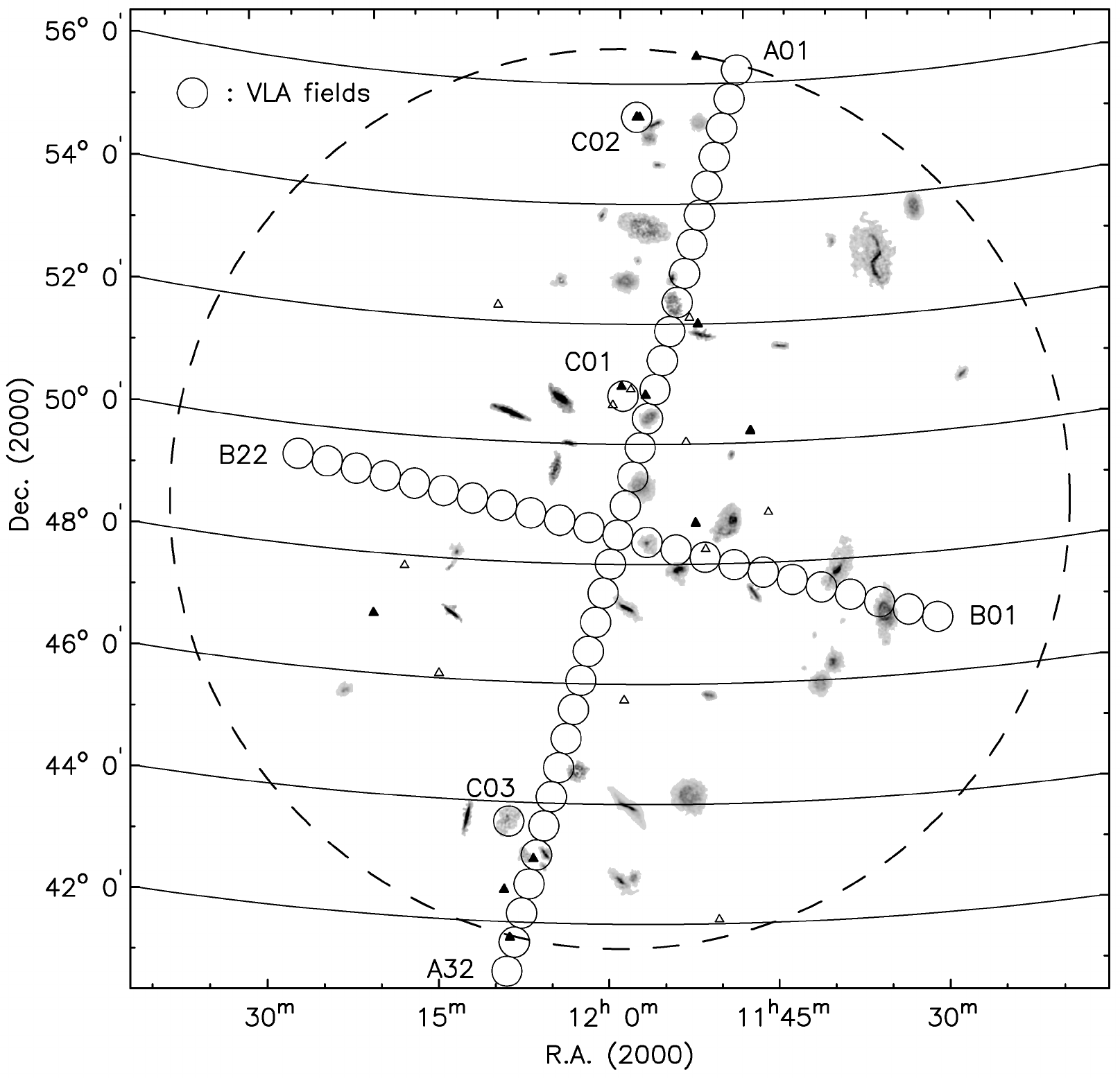} 
\caption{The layout of the VLA pointings in the Ursa Major cluster. The cross-pattern of circles indicate the VLA pointing centers and the size of the FWHM of the primary beam. Greyscale images show the total HI maps of 56 optically selected and known cluster members, previously obtained with the WSRT. the triangles indicate the positions of the remaining 23 cluster members. Solid triangles; galaxies brighter than the optical completion limit but without sufficient HI for useful synthesis imaging. These are mainly lenticulars or S0a galaxies whose cluster membership was established through optical redshifts. Open triangles; dwarf galaxies fainter than the optical completion limit. Nearly all of these have measured global HI profiles from single dish observations. The large dashed circle shows the 15 degree diameter spatial window on the sky. The solid curved horizontal lines run along constant declination.}\label{fig:UMaVLA}
\end{figure*}

\section{The VLA blind HI survey}

\subsection{Observational setup}
A blind VLA D-array HI survey was designed to measure the slope at the low-mass end of the UMa HIMF. If 16 per cent of the volume is observed, the overdensity of the nearby spiral-rich UMa volume provides a sufficient number of detections per HI mass-bin down to 10$^7$M$_\odot$ to measure the slope of the HIMF. The 45 arcsec synthesized beam of the Karl G. Jansky Very Large Array (VLA) in its D-configuration corresponds to 3.8 kpc at a distance of 17.4 Mpc. The observations were performed over ten 8-hour runs between March 11 and May 22 of 1999.

The lay-out of the blind survey is a cross-pattern shown in Figure \ref{fig:UMaVLA}. This cross-pattern contains two orthogonal rows of respectively 32 and 22 pointings, separated by 32 arcmin, the FWHM of the primary beam.  After mosaicing the individual pointings we cover a larger volume than just taking the FWHM of the primary beam for an individual pointing into account. The total observed volume is 20 Mpc$^3$ (16\% of the total UMa volume) at an HI sensitivity of $\sim$ $5 \times 10^7$M$_\odot$. For higher HI masses the volume is somewhat larger as we can detect galaxies with a higher HI mass well beyond the FWHM of the mosaiced primary beam pattern. The major axis of this cross runs along supergalactic longitude and through the middle of the region. Three additional pointings were observed, aimed at the two brightest lenticulars NGC 3998 and NGC 4026, and the galaxy NGC 4138. Results from these three targeted observations will be ignored in the rest of this paper. 

The greyscale images in Figure \ref{fig:UMaVLA} show the total HI maps of 56 UMa galaxies, individually enlarged by a factor four, as imaged by the WSRT.  Several of these galaxies are located closely together and their HI images are adjacent or overlapping. Detailed information on most of these galaxies are presented in \citet{verheijen2001b}. As can be seen in Figure \ref{fig:UMaVLA}, the blind VLA-D survey samples both crowded and empty regions of the volume. Indeed, the distribution of galaxies within the UMa volume is not uniform.

For the VLA observations, a bandwidth of 3.125 MHz with 63 channels ($\Delta V = 5.15$ km s$^{-1}$) is sufficient to cover the volume's velocity window, since the velocity dispersion of the known members is low. Given this bandwidth and a dual polarization setup, the VLA correlator provided a velocity resolution of 10.3 km/s after offline Hanning smoothing. The typical integration time per pointing was 70 minutes.  

\begin{figure}
\includegraphics[width=0.5\textwidth]{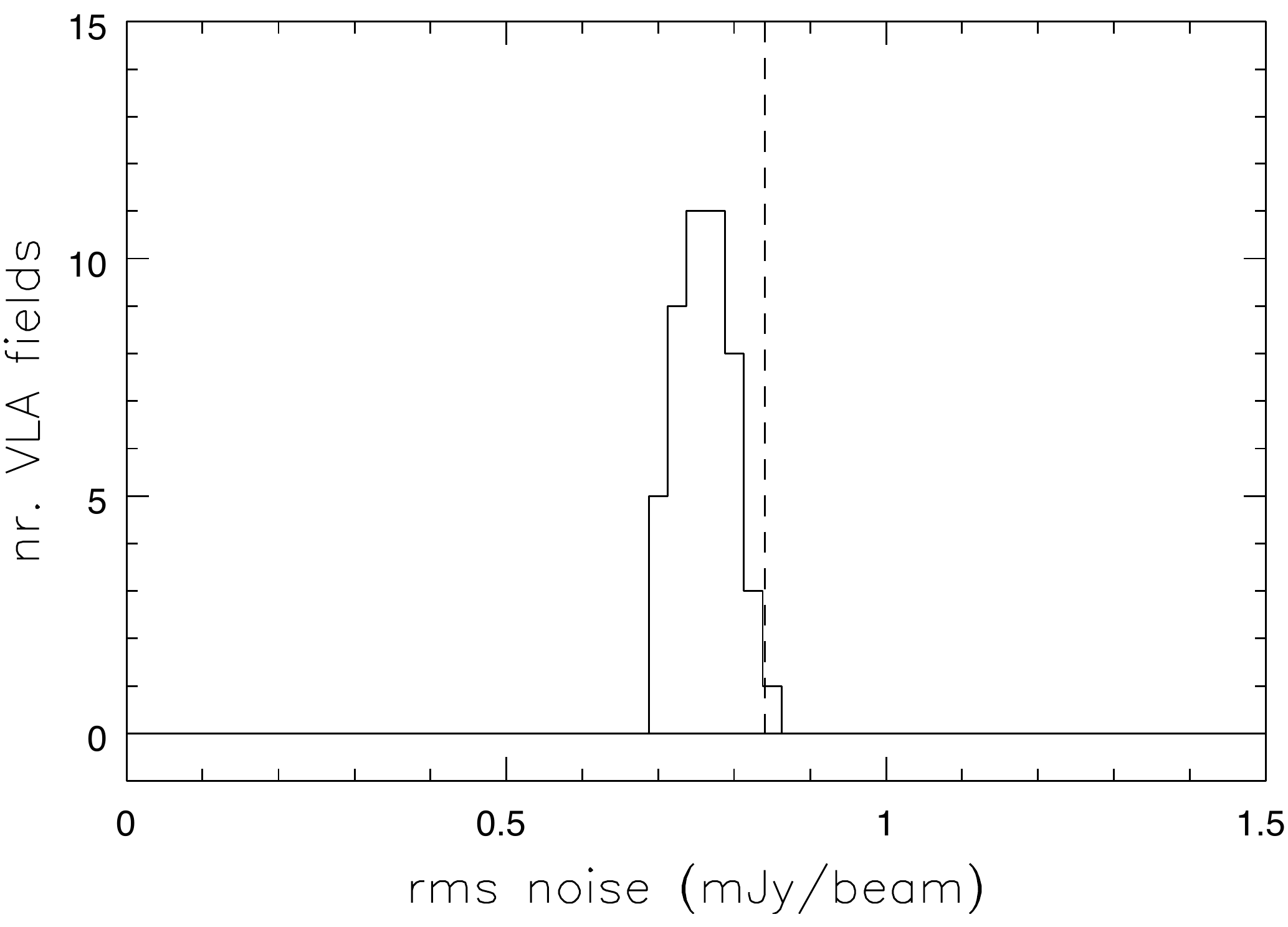}
 \caption{Distribution of the rms noise levels in the data cubes of the 54 individual VLA pointings. The narrow distribution indicates a uniform sensitivity over the entire survey. The dashed line indicates the noise level of 0.84 mJy/beam adopted for the entire area.}\label{fig:rms}
\end{figure}

The blind VLA survey was supplemented with a wide-field R-band imaging survey on the CFHT of the cross-pattern of the VLA fields. This optical survey was aimed at measuring the faint end of the LF while it would provide optical morphologies and luminosities of the HI detected dwarfs. In return, the HI detections would provide redshifts for the optically faint yet gas-rich dwarf galaxies in the UMa volume. The LF of the CFHT sample has the parameters $M^{*}_{R} = -21.44$ and $\alpha = -1.01$ \citep{trentham2001} which is flat compared to the LF of surveys of other volumes, for example the LF of the SDSS sample with a slope of $-1.20 \pm 0.03$ \citep{blanton2001}.

\subsection{Data reduction and analysis}

The collected visibility data were flagged and calibrated in the standard VLA fashion with the Astronomical Image Processing System (AIPS) \citep{greisen2003}. Every 40 minutes, nearby 3C295 was observed for flux and phase calibration and to monitor the time variable `3-MHz ripple' in the bandpass. The complex gain corrections and bandpass shapes were linearly interpolated in time between the calibration observations.  Because the observations were carried out within the protected 21cm-band, only minimal editing of the data was required to remove Radio Frequency Interference. To verify the quality of the data, the flagged and calibrated visibilities of each pointing were Fourier transformed using the IMAGR task in AIPS, applying a {\em Robust} weighting of 1. The data cubes were then imported into the Groningen Image Processing SYstem (GIPSY) software package (\citet{vanderhulst1992}; \citet{vogelaar2001}) for further inspection. After spectral Hanning smoothing, continuum subtraction and a standard H\"{o}gbom cleaning \citep{hogbom1974} of the data cubes, the noise was inspected and measured in each cube. Figure \ref{fig:rms} shows the distribution of the achieved rms noise in the 54 individual data cubes. The average noise level is S$_\mu$=0.76 mJy beam$^{-1}$ with a variance of S$_\sigma$=0.04 mJy beam$^{-1}$, demonstrating that a rather uniform noise was achieved over the entire surveyed area.

After data calibration and quality assessment with AIPS and GIPSY, the Miriad software package \citep{sault1995} was used to construct image mosaics in order to increase the survey sensitivity in between adjacent VLA pointings. Due to limited computing resources at the time these data were reduced, 
the survey cross-pattern was chopped up into nine sub-mosaics; four mosaics along the major axis, four mosaics along the minor axis, and a mosaic of the central five pointings. Each mosaic cube has non-blank pixels where the VLA primary beam correction is less than a factor 43. The mosaic algorithm weighs the visibility data of each pointing by the theoretical noise, producing an image cube with near-uniform noise. 

Since we do not know where in our blind survey volume we may find HI emission, we first used the Miriad task 'invert' to produce un-cleaned mosaic image cubes in which we searched for obvious HI emission. Using GIPSY tasks, the continuum sources and their instrumental responses were removed by iteratively fitting a linear baseline to the spectrum at each pixel. After each baseline fit, data points above and below a certain noise-related clip level were identified and excluded from a subsequent new baseline fit. The clip levels were gradually decreased to $\pm2.5\sigma$ during this iterative process. In this way, any unknown low-level HI emission is retained above the zero-level baseline. We visually identified any obvious HI emission in these continuum-subtracted yet un-cleaned cubes and manually constructed 3-dimensional masks that enclose the HI signal. Subsequently, we used the Miriad task 'mossdi' to clean the mosaic cubes using the \citet{sdi1984} cleaning algorithm, guided by the masks produced in the previous step. The resulting mosaic image cubes are free from continuum sources and side-lobes from obvious HI emission and can be searched more objectively for fainter HI sources that were missed when visually identifying the obvious sources of HI emission. Finally, for each of the nine mosaics, the Miriad task 'mossen' was used to construct an adjusted gain maps describing the effective primary beam attenuation across the mosaics.

\begin{figure}
\includegraphics[width=0.5\textwidth]{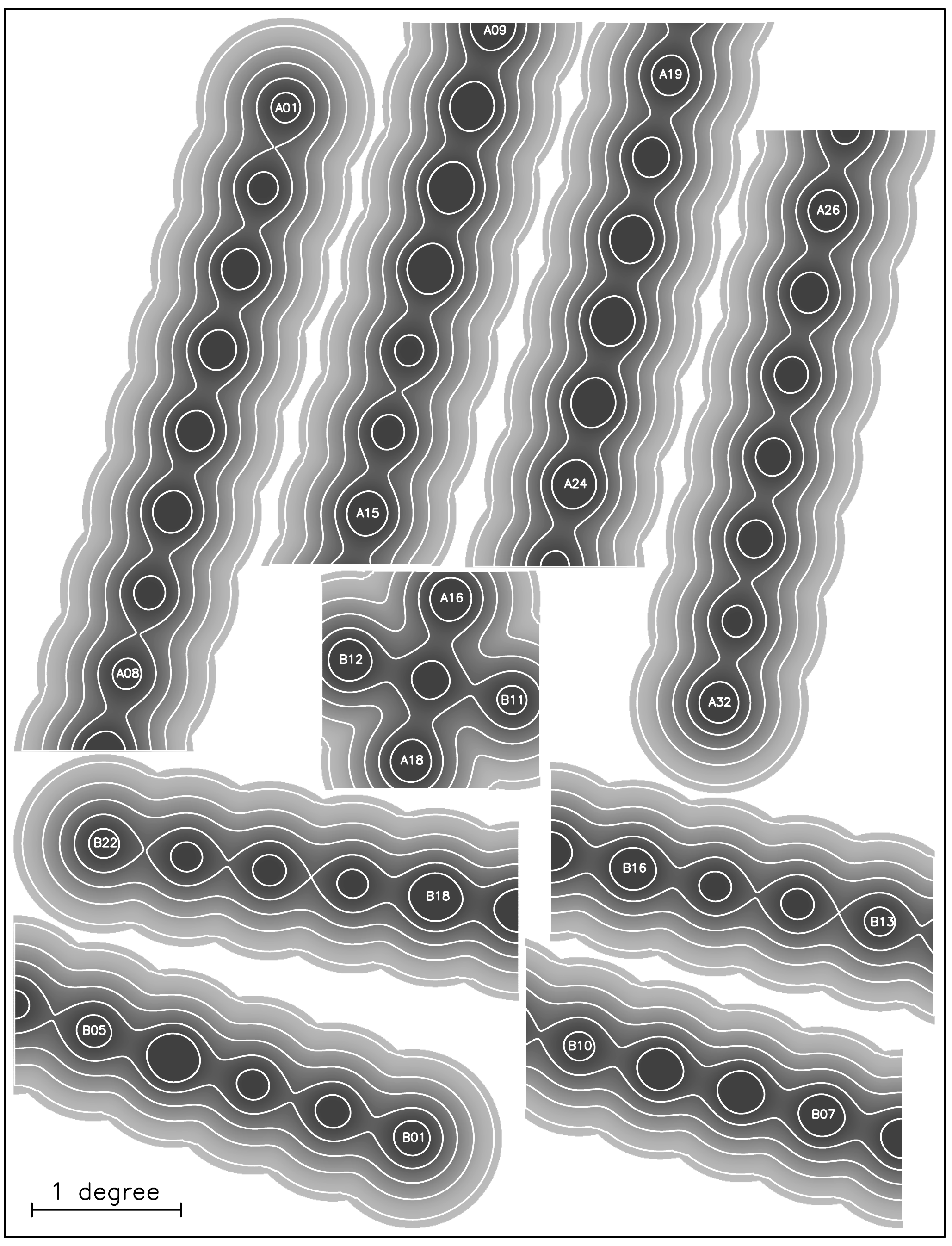}
\caption{Adjusted gain patterns after combining the nine sub-mosaics into a single data cube. Contour levels indicating a normalised gain of 0.95, 0.75, 0.50, 0.25, and 0.05.}\label{fig:pbp}
\end{figure}

To reduce the number of blank pixels to be processed during the analysis, the nine mosaic cubes were cropped and compiled into a single 'master' data cube whereby overlapping regions covered by adjacent mosaics were removed. This was done for both the nine mosaic image cubes and the nine adjusted gain cubes. Figure \ref{fig:pbp} shows a single channel from the 'master' adjusted gain cube in which the white areas correspond to blank pixels. Obviously, no meaningful coordinate system is attached to this cube as it merely serves as input for our source finding. The total number of non-blank pixels in each channel of this 'master' cube is 1.609.322 and with 15"$\times$15" pixels, our mosaic covers a total of 27.9 deg$^2$ or 15.8\% of the sky area enclosed by the dashed circle in Figure \ref{fig:UMaVLA}.

To improve sensitivity to line emission profiles that are broader than the instrumental velocity resolution, the already Hanning smoothed 'master' data cube at a velocity resolution of 10.3 km s$^{-1}$ (R2) was smoothed further in velocity using a Gaussian smoothing kernel to obtain velocity resolutions of 4, 6, 8 and 10 channels, corresponding to 20.6, 30.9, 41.2 and 51.5 km s$^{-1}$ respectively (indicated with R4, R6, R8 and R10 hereafter). To improve sensitivity to extended HI emission of lower column density, the data were also smoothed spatially from a resolution of 45" to Gaussian synthesised beams with a FWHM of 60", 90", 120", 150" and 180". These 30 smoothed data cubes form our basic observational material. Table \ref{tab:rmsmhi} lists the mean rms noise level for each of the 30 resolutions, as well as the corresponding HI masses and HI column densities detectable in a single Gaussian resolution element with a peak flux at the 6$\sigma$ level at the centre of the primary beam and at the average distance of 17.4 Mpc. 

For a given velocity and angular resolution, the minimum detectable HI mass varies by a factor 4.3 for a dwarf galaxy located near the centre of the primary beam and at the near-side of UMa volume (14.7 Mpc) compared to a dwarf galaxy near the half-power point of the primary beam  at the far-side of the UMa volume (21.6 Mpc). The column density sensitivity, on the other hand, is almost independent over the distance range of the UMa volume and is only affected by the primary beam attenuation. The smallest, spatially unresolved, low-mass HI cloud, supported only by turbulent motion with a typical velocity dispersion of 8 km s$^{-1}$ resulting in a Gaussian profile with a FWHM of $\sim$20 km s$^{-1}$, located near the centre of the primary beam and at the near-side of the UMa volume, would have a minimum detectable HI mass of 3.5$\times$10$^6$ M$_\odot$ at the 6$\sigma$-level. The volume within which such a small HI mass can be detected in our survey, however, is negligible.

\begin{figure}
\includegraphics[width=0.5\textwidth]{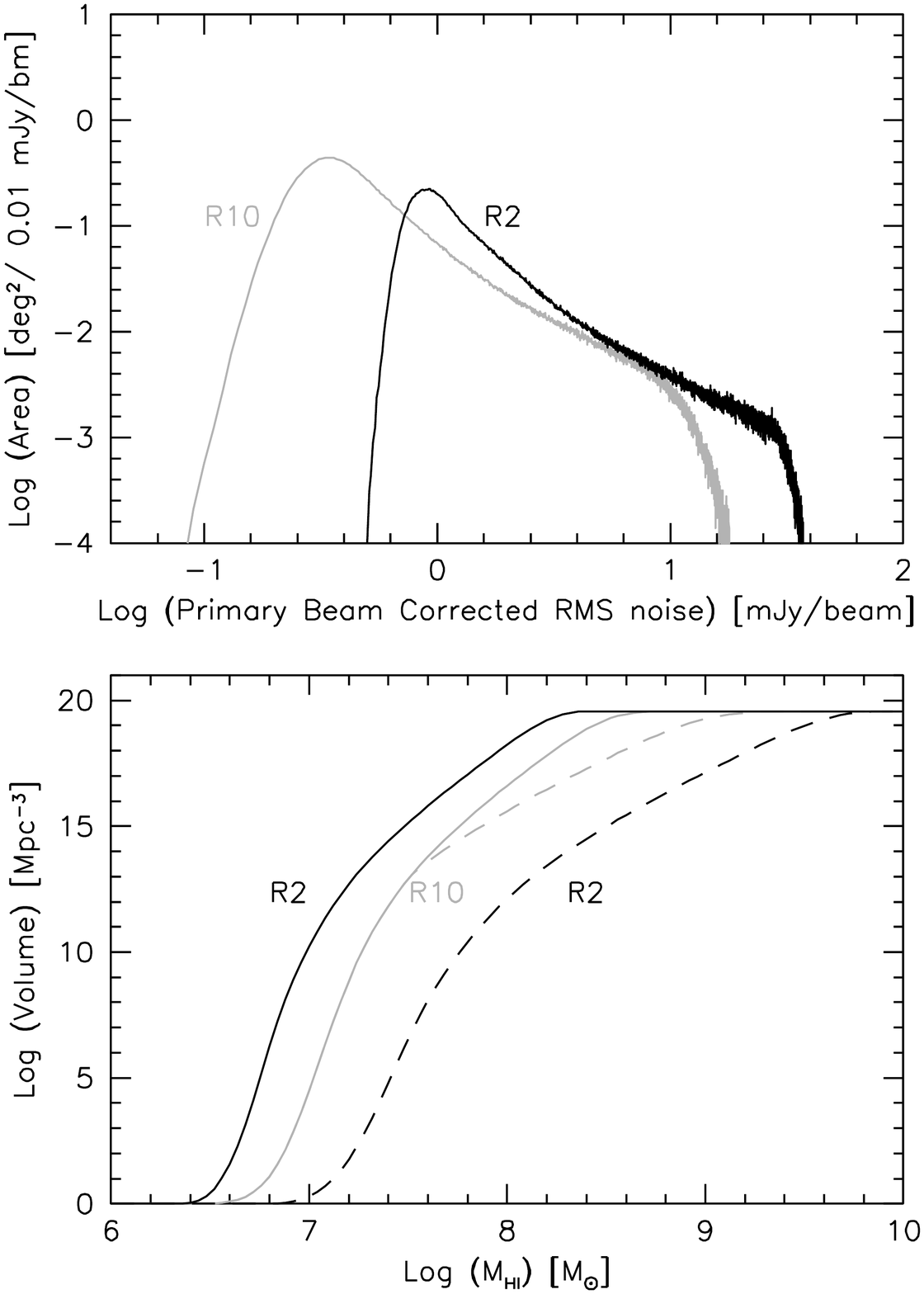}
 \caption{Upper panel: Surveyed area on the sky as a function of primary beam corrected noise for 2 different resolutions (R2 and R10). Lower panel: Volume sensitivity as a function of HI mass for different spectral resolutions (R2 and R10) and linewidth assumptions (solid line: W$_{20}$ = FWHM and dashed line: W$_{20}$ = 0.16 M$_{HI}^{1/3}$) \citep{zwaan1997}.}\label{fig:pbnoise}
\end{figure}

\begin{table}
\caption{RMS noise levels and 6$\sigma$ detection limits for the various spatial and velocity resolutions.}\label{tab:rmsmhi}
\begin{tabular}{rrrrrr}
Res.                 &  R2   &  R4   &  R6   &  R8   &  R10  \\
                     & 10.3  & 20.6  & 30.9  & 41.2  & 51.5  \\
 & km s$^{-1}$ & km s$^{-1}$ & km s$^{-1}$ & km s$^{-1}$ & km s$^{-1}$ \\
\hline \hline
\multicolumn{6}{l}{\underline{RMS noise levels (mJy/beam)}}  \\
45$^{\prime\prime}$  &  0.79 &  0.52 &  0.42 &  0.35 &  0.31 \\
60$^{\prime\prime}$  &  0.94 &  0.62 &  0.50 &  0.43 &  0.37 \\
90$^{\prime\prime}$  &  1.5  &  0.98 &  0.78 &  0.67 &  0.59 \\
120$^{\prime\prime}$ &  2.0  &  1.3  &  1.1  &  0.96 &  0.82 \\
150$^{\prime\prime}$ &  2.5  &  1.7  &  1.4  &  1.2  &  1.1  \\
180$^{\prime\prime}$ &  3.1  &  2.1  &  1.7  &  1.5  &  1.3  \\
\multicolumn{6}{l}{\underline{6$\sigma$ HI mass limit (10$^6$ M$_\odot$)}} \\
45$^{\prime\prime}$  &  3.7  &  4.9  &  5.9  &  6.6  &  7.2  \\
60$^{\prime\prime}$  &  4.4  &  5.8  &  7.0  &  8.1  &  8.6  \\
90$^{\prime\prime}$  &  7.0  &  9.2  & 11    & 13    & 14    \\
120$^{\prime\prime}$ &  9.4  & 12    & 16    & 18    & 19    \\
150$^{\prime\prime}$ & 12    & 16    & 20    & 23    & 26    \\
180$^{\prime\prime}$ & 15    & 20    & 24    & 28    & 30    \\
\multicolumn{6}{l}{\underline{6$\sigma$ HI column density limit (10$^{19}$ cm$^{-2}$)}} \\
45$^{\prime\prime}$  &  2.9  &  3.8  &  4.7  &  5.2  &  5.7  \\
60$^{\prime\prime}$  &  2.0  &  2.6  &  3.1  &  3.6  &  3.8  \\
90$^{\prime\prime}$  &  1.4  &  1.8  &  2.2  &  2.5  &  2.7  \\
120$^{\prime\prime}$ &  1.0  &  1.4  &  1.7  &  2.0  &  2.1  \\
150$^{\prime\prime}$ &  0.83 &  1.1  &  1.4  &  1.6  &  1.8  \\
180$^{\prime\prime}$ &  0.72 &  1.0  &  1.2  &  1.4  &  1.5  \\
\end{tabular}     
\end{table}

\section{Observational results}\label{sec:obsres}

In each of the 30 smoothed mosaic data cubes a $\pm$6$\sigma$ clip was applied to find the HI emission and possibly HI absorption against the many continuum sources. This clipping process yielded 67 unique detections including 26 known galaxies, 16 bonafide new HI sources, 11 doubtful emission line objects, 7 objects with negative amplitudes and 7 'objects' related to obvious imaging artefacts. Each of these detection categories will now be briefly discussed.

The 26 previously catalogued galaxies with known redshifts are listed in Table \ref{tab:HIdet}. In the imaged area, 25 known members are located. All of them were detected with the exception of NGC 4143, which is an HI poor S0/a system located at the very edge of the imaged area where the primary beam response is less than five per cent. Detected non-members are the well-known galaxies NGC 3850 and NGC 3898 ($v_{\rm sys} > 1210$ km s$^{-1}$) in the northern part of the survey area. Many of these known galaxies are bright and have broad HI emission profiles. The velocity wings of several of these galaxies fall outside the observed velocity range and the HI properties of those galaxies as listed in Table \ref{tab:HIdet} were taken from \citet{verheijen2001b} or elsewhere from the literature.

A total of sixteen previously unknown HI emission line objects were detected, all with optical counterparts that are cataloged by the NASA Extra-galactic Database (NED). These objects are listed in Table \ref{tab:newHIdet} and all of these objects have optical counterparts in SDSS images. Most new detections have integrated HI fluxes below $\int SdV \sim 3.30$ Jy km s$^{-1}$, except for NGC 4111 (see Table \ref{tab:newHIdet}) which has a total integrated HI flux of $\int SdV = 12.26$ Jy km s$^{-1}$ but was not detected in HI before our observations were carried out. All XV-diagrams and images of the new HI detections are shown in Figures \ref{fig:6a} and \ref{fig:6b} of the appendix. The detailed morphologies and other characteristics of these new detections will be discussed in a forthcoming paper.

The 11+7 doubtful emission and absorption line objects are questionable for several reasons. Concerning the seven `absorption' profiles, six of them occur at positions without a continuum source while the seventh negative object occurs in a single pixel on the flank of a 13 mJy continuum point source and at the very edge of the bandpass. This suggests that all negative objects are actually negative noise peaks. Concerning the eleven doubtful emission profiles, six of them occur in a single pointing with a slightly higher noise level, with a peak flux <6.5-$\sigma$ and only at the lowest velocity resolutions. The other five doubtful emission line objects balance the negative noise peaks in numbers and are therefore statistically likely to be positive noise peaks. 

Finally, there are seven `detections' related to obvious imaging artefacts due to residuals in the spectra of bright continuum sources caused by imperfect bandpass corrections, due to the side-lobes of a bright HI source outside the imaged field-of-view (NGC 3726), or due to ripples in the noise caused by RFI residuals that become apparent only at the lowest angular resolutions and that were not identified through visual inspection at the early stages of data reduction.

\begin{table*}
\begin{sideways}
\begin{minipage}[c]{230mm}       
\caption{HI detected galaxies, previously catalogued and known to contain HI.}\label{tab:HIdet}
\begin{tabular}{lccrrrrrrrrrccccc}
 \multicolumn{5}{l}{ } & \multicolumn{7}{|c|}{Values from literature} & \multicolumn{5}{c}{Values from this work}\\
 & Name & Field & R.A. & Dec. & \multicolumn{1}{|r}{M(R)} & W$_{\rm 20}$ & R & V$_{\rm hel}$ & V$_{\rm sys}$ & $\int$Sdv & \multicolumn{1}{c|}{Ref} & W$_{\rm 20}$ & V$_{\rm hel}$ & V$_{\rm sys}$ & $\int$Sdv  & Comments\\
 & & & h m s & d m s & \multicolumn{1}{|r}{mag} & km s$^{-1}$ & km s$^{-1}$ & km s$^{-1}$ & km s$^{-1}$ & Jy km s$^{-1}$ &  \multicolumn{1}{c|}{ } & km s$^{-1}$ & km s$^{-1}$ & km s$^{-1}$ & Jy km s$^{-1}$ & \\
 \hline \hline
 \multicolumn{17}{l}{\underline{Inside Ursa Major volume}} \\
 & NGC 3726 & B02 & 11:33:21.2 & 47:01:45 & -21.53 & 287 & 10.3 &  866 &  913 & 90   & 1 &  283 &   860 &   919 &  96.4  &                                \\
 & NGC 3769 & B03 & 11:37:44.1 & 47:53:33 & -20.14 & 265 & 10.3 &  737 &  796 & 62   & 1 & $>$118 &  $<$794 &  $<$853 &  26.6  & edge FOV               \\
 & 1135+48  & B03 & 11:37:50.5 & 47:52:54 & -17.35 & 123 & 10.3 &  798 &  857 &  6.6 & 1 &  &       &       &        & confused                       \\
 & NGC 3877 & B07 & 11:46:07.8 & 47:29:41 & -21.53 & 373 & 10.3 &  895 &  955 & 20   & 1 & 352 &   892 &   952 &  18.8  &                                \\
 & NGC 3906 & B08 & 11:49:40.1 & 48:25:34 & -18.65 &  58 &  7.4 &  959 & 1024 &  5.1 & 2 & 42 &   961 &  1026 &   3.5  &                                \\
 & NGC 3913 & A04 & 11:50:38.8 & 55:21:13 & -18.96 &  62 &  7.4 &  952 & 1049 & 11   & 2 & 52 &   949 &  1046 &  13.9  &                                \\
 & UGC 6840 & A09 & 11:52:06.9 & 52:06:29 & -17.94 & 157 &  7.4 & 1035 & 1118 & 13   & 2 & 154 &  1017 &  1100 &  16.8  &                                \\
 & NGC 3949 & B10 & 11:53:41.6 & 47:51:33 & -20.79 & 287 & 10.3 &  800 &  864 & 45   & 1 & 271 &   799 &   863 &  34.3  &                                \\
 & NGC 3953 & A09 & 11:53:48.9 & 52:19:37 & -22.01 & 442 & 10.3 & 1052 & 1137 & 40   & 1 & $>$338 & $>$1006 & $>$1091 & $>$34.8  & edge band               \\
 & UGC 6917 & A13 & 11:56:28.5 & 50:25:45 & -19.30 & 209 & 10.3 &  911 &  988 & 26   & 1 & 199 &   910 &   987 &  25.8  &                                \\
 & NGC 3985 & B11 & 11:56:41.6 & 48:20:06 & -19.14 & 160 & 10.3 &  948 & 1016 & 16   & 1 & 160 &   949 &  1017 &  15.5  &                                \\
 & UGC 6922 & A12 & 11:56:52.0 & 50:48:59 & -17.63 & 159 &  7.4 &  893 &  970 &  8.9 & 2 & 140 &   891 &   968 &   9.9  &                                \\
 & UGC 6930 & A15 & 11:57:17.3 & 49:16:59 & -19.61 & 137 & 10.3 &  777 &  849 & 43   & 1 & 135 &   776 &   848 &  38.2  &                                \\
 & UGC 6956 & A12 & 11:58:25.5 & 50:55:02 & -17.47 &  67 &  7.4 &  918 &  998 & 11   & 2 & 60 &   914 &   994 &   9.2  &                                \\
 & NGC 4010 & A20 & 11:58:34.7 & 47:15:34 & -19.79 & 278 & 10.3 &  902 &  965 & 38   & 1 & $>$221 & $<$920 &  $<$947 & $>$22.4  & edge FOV              \\
 & NGC 4051 & A25 & 12:03:09.5 & 44:31:54 & -21.58 & 255 & 10.3 &  700 &  753 & 36   & 1 & $>$249 &  $<$706 &  $<$747 & $>$34.4  & edge band               \\
 & UGC 7089 & A28 & 12:05:57.8 & 43:08:36 & -18.89 & 157 & 10.3 &  770 &  818 & 17   & 1 & 156 &   771 &   819 &  16.3  &                                \\
 & 1203+43  & A28 & 12:05:59.2 & 42:54:10 & -15.42 &  74 & 33.3 &  756 &  803 &  0.7 & 3 & 40 &   762 &   809 &   0.71 &                                \\
 & UGC 7094 & A28 & 12:06:10.8 & 42:57:23 & -17.69 &  84 & 10.3 &  780 &  827 &  2.9 & 1 & 101 &   780 &   827 &   2.6  &                                \\
 & NGC 4117 & A28 & 12:07:46.1 & 43:07:35 & -19.00 & 289 & 10.3 &  934 &  982 &  6.9 & 1 & 256 &   947 &   995 &   2.7  &                                \\
 & NGC 4118 & A28 & 12:07:52.8 & 43:06:42 & -16.43 & 128 & 33.3 &  661 &  709 &  1.4 & 3 & 49 &   640 &   688 &   0.34 &                                \\
 & UGC 7129 & A31 & 12:08:55.3 & 41:44:27 & -18.52 & 160 & 10   &  947 &  990 &  4.7 & 4 & 130 &   933 &   976 &   0.70 &                                \\
 \multicolumn{17}{l}{\underline{Outside Ursa Major volume}} \\
 & NGC 3850 & A02 & 11:45:35.6 & 55:53:13 &   $-$    & 173 &  7.4 & 1149 & 1226 & 12   & 2 & $>$117 & $>$1116 & $>$1193 &  $>$5.8  & edge band \\
 & NGC 3898 & A01 & 11:49:15.2 & 56:05:04 &   $-$     & 482 &  7.4 & 1171 & 1271 & 34   & 2 & $>$53 &  $>$965 & $>$1065 &  $>$5.6  & edge FOV \\
\end{tabular}
\begin{list}{}{}
\item[(1)] \citet{verheijen2001a}
\item[(2)] \citet{appleton1982}
\item[(3)] \citet{vanderburg1987}
\item[(4)] \citet{bottinelli1999}
\end{list}
\end{minipage}
\end{sideways}
\end{table*}

\begin{table*}
\begin{sideways}
\begin{minipage}[c]{230mm}       
\caption{HI detected galaxies, previously uncatalogued.}\label{tab:newHIdet}
\begin{tabular}{llcccrrclrrrrrl}
 
 & Name & Field & R.A. & Dec. & Rad & P.A. & eps & Res & S/N & W$_{20}$ & V$_{hel}$ & V$_{sys}$ & $\int$Sdv  & SDSS name \\
 & & & h m s & d m s & '' & deg & & & & km s$^{-1}$ & km s$^{-1}$ & km s$^{-1}$ & Jy km s$^{-1}$ & \\
\hline \hline 
\multicolumn{15}{l}{\underline{Inside Ursa Major volume}} \\
 & UMa-HI01 & A02 & 11:46:34.2 & 55:49:16 &  15.8 & 171 & 0.293 & R2-45  & 121   &  81 & 1076 & 1172 &  0.98 &  J114634.06+554917.0\\
 & UMa-HI02 & B08 & 11:46:49.7 & 48:05:30 &   9.9 & 170 & 0.528 & R4-45  &  14   &  45 & 1039 & 1102 &  0.22 &  J114649.67+480530.5 \\
 & UMa-HI03 & A01 & 11:48:20.3 & 56:20:45 &  13.5 &  45 & 0.190 & R6-60  &  16   &  35 & 1078 & 1179 &  1.25 &  J114820.16+562045.8\\
 & MRK 1460 & B09 & 11:50:50.0 & 48:15:05 &   9   &  31 & 0.293 & R8-45  &  22   &  67 &  821 &  886 &  0.25 &  J115050.04+481505.2 \\
 & UMa-HI04 & B09 & 11:50:59.8 & 47:57:47 &  10.4 & 146 & 0.358 & R6-45  &   6.2 &  18 &  950 & 1014 &  0.06 &  J115059.60+475749.5\\
 & UMa-HI05 & A07 & 11:51:53.7 & 53:06:00 &  12.2 &  24 & 0.007 & R4-45  &  21   &  40 & 1117 & 1204 &  0.21 &  J115153.66+530558.2\\
 & UMa-HI06 & B10 & 11:53:11.3 & 48:11:19 &   8.9 &   0 & 0.247 & R2-45  &  48   &  33 &  756 &  821 &  0.31 &  J115311.13+481118.5 \\
 & UMa-HI08 & A16 & 11:58:11.9 & 48:52:50 &  25.6 & 162 & 0.288 & R2-45  & 254   &  91 &  819 &  890 &  2.36 &  J115811.54+485253.1\\
 & UMa-HI09 & A16 & 11:58:26.0 & 48:57:36 &   9.0 & 156 & 0.444 & R2-45  &  32   &  28 &  956 & 1027 &  0.22 &  J115825.88+485737.1\\
 & UMa-HI10 & A15 & 11:59:57.9 & 49:33:53 &  23.5 &  47 & 0.302 & R4-60  &  40   &  77 & 1135 & 1209 &  1.14 &  J115957.67+493349.8\\
 & UMa-HI11 & A18 & 12:00:35.0 & 47:46:22 &  39.7 & 106 & 0.300 & R2-60  & 278   &  95 &  658 &  725 &  3.23 &  J120035.41+474626.1\\
 & UMa-HI12 & A24 & 12:02:43.7 & 45:11:27 &  31.4 &  57 & 0.733 & R10-45 &   6.6 & 152 &  714 &  770 &  0.77 &  J120243.72+451127.9\\
 & UMa-HI13 & A28 & 12:05:50.1 & 42:54:56 &  15.4 &  44 & 0.242 & R4-45  &  15   &  43 &  772 &  819 &  0.46 &  J120550.87+425509.9\\
 & NGC 4111 & A28 & 12:07:03.1 & 43:03:58 & 134   & 150 & 0.775 & R6-180 &  60   & 244 &  813 &  861 & 12.26 &  J120703.12+430356.4 \\
 & UMa-HI14 & A29 & 12:07:03.1 & 42:42:50 &       &     &       & R2-60  &  42   &  61 &  879 &  925 &  0.84 &  no optical counterpart\\
\multicolumn{15}{l}{\underline{Outside Ursa Major volume}} \\
 & UMa-HI07 & A11 & 11:56:16.4 & 51:17:06 &   8.6 & 174 & 0.230 & R2-45  &  13   &  28 & 1153 & 1234 &  0.15 &  J115616.24+511706.9\\
 & UMa-HI15 & A31 & 12:07:51.6 & 41:33:47 &   8.3 &  48 & 0.196 & R2-45  &  43   &  37 & 1141 & 1182 &  0.30 &  J120751.57+413346.9\\
 & UMa-HI16 & A32 & 12:08:24.4 & 41:24:07 &  16.6 & 123 & 0.596 & R4-45  &  31   &  46 & 1012 & 1053 &  0.35 &  J120824.52+412404.8\\

\multicolumn{15}{l}{\underline{New $<$6-sigma HI detections with optical counterpart}} \\

 & UMa-HI17 & A01 & 11:45:47.6 & 56:26:43 &   9.3 &  20 & 0.471 & R2-45  &   5.9 &  60 & 1123 & 1223 &  0.35 &  J114547.46+562642.8\\
\end{tabular}
\end{minipage}
\end{sideways}
\end{table*}

\section{Volume corrections}\label{sec:volcor1}

To properly determine the HIMF in the UMa volume, it is necessary to correct for the low-mass galaxies that could not be detected because they are located at the far end of the survey volume, or towards the edges of the imaged area where the primary beam attenuation is significant, or in survey areas where the observational noise is elevated. We need to understand these effects to be able to calculate the correct volume corrections and note that different assessments of the volume corrections might lead to different slopes of the HIMF (\citet{rosenberg2002}, \citet{zwaan2005}). To make an initial assessment of the severity of the incompleteness due to a varying sensitivity across the UMa survey volume, we illustrate two characteristics of our survey. 

First, we divide the R2 and R10 'master' mosaic cubes at the highest angular resolution by the mosaicked relative gain cube (Fig. \ref{fig:pbp}) in order to apply the effective primary beam correction. As a consequence, the noise in the resulting cube varies strongly across the survey area. At each pixel position in the cube we calculate the rms noise in the corresponding spectrum and construct a 2-dimensional map of the spectral noise. From this spectral noise map we calculate the number of pixels and corresponding survey areas in bins of 0.01 mJy beam$^{-1}$ and plot this area as a function of rms noise in the upper panel of Figure \ref{fig:pbnoise}. The noise varies by 1.7 dex (close to the expected factor 43) while most of the survey area is covered by lower noise values.

Second, we consider the maximum volume $V_{\rm max}$ in which a galaxy of a certain HI mass can be detected at the 6$\sigma$ level, taking into account the varying noise across the survey area after primary beam correction. For each position ${\vec x}$ on the sky, we calculate the maximum distance $D_{\rm max}({\vec x})$ at which a particular HI mass can be detected given the spectral noise at that position. Note that $D_{\rm max}({\vec x})$ is restricted to the interval 14.7$-$21.6 Mpc. Subsequently we calculate the volume element $V({\vec x})$ between 14.7 Mpc and $D_{\rm max}({\vec x})$, covered by a pixel at that sky position. We then add all volume elements of all the pixels to calculate the total maximum volume $V_{\rm max}$ covered by the survey area in which a particular HI mass can be detected. The results are plotted in the lower panel of Figure \ref{fig:pbnoise} for two assumptions. The solid lines assume that the HI flux is contained within a single Gaussian resolution element. The dashed lines adopt the empirical relation between HI mass and line width as established by \citet{zwaan1997}, also assuming that the HI profile is Gaussian and that the minimum line width is set by the velocity resolution of our observations. Considering the solid black line and corresponding assumptions, we conclude that HI masses larger than 10$^{8.4}$ can be detected throughout the entire survey volume while there is no volume within which HI masses less than 10$^{6.4}$ can be detected.

For our UMa survey, the $1/V_{\rm max}$ method \citep{schmidt1968} is most appropriate. The essence of this method is to determine for every object in our sample the maximum volume in which this particular object could have been detected, as discussed above and illustrated in the bottom panel of Figure \ref{fig:pbnoise}. A concern often raised is that the $1/V_{\rm max}$ method can be biased by a non-uniform distribution of galaxies in the survey volume. As can be seen from Figure \ref{fig:UMaVLA}, indeed, the UMa galaxies are not uniformly distributed. However, according to \cite{zwaan1997}, the $1/V_{\rm max}$ method is nevertheless the best method for a small sample like ours, since maximum likelihood methods can produce large errors when only a few galaxies per bin are present. We also note that the effect of a non-uniform distribution is somewhat mitigated by the fact that our survey area covers both high and low density regions within the UMa volume.

\begin{figure}
 \centerfloat
\includegraphics[width=0.5\textwidth]{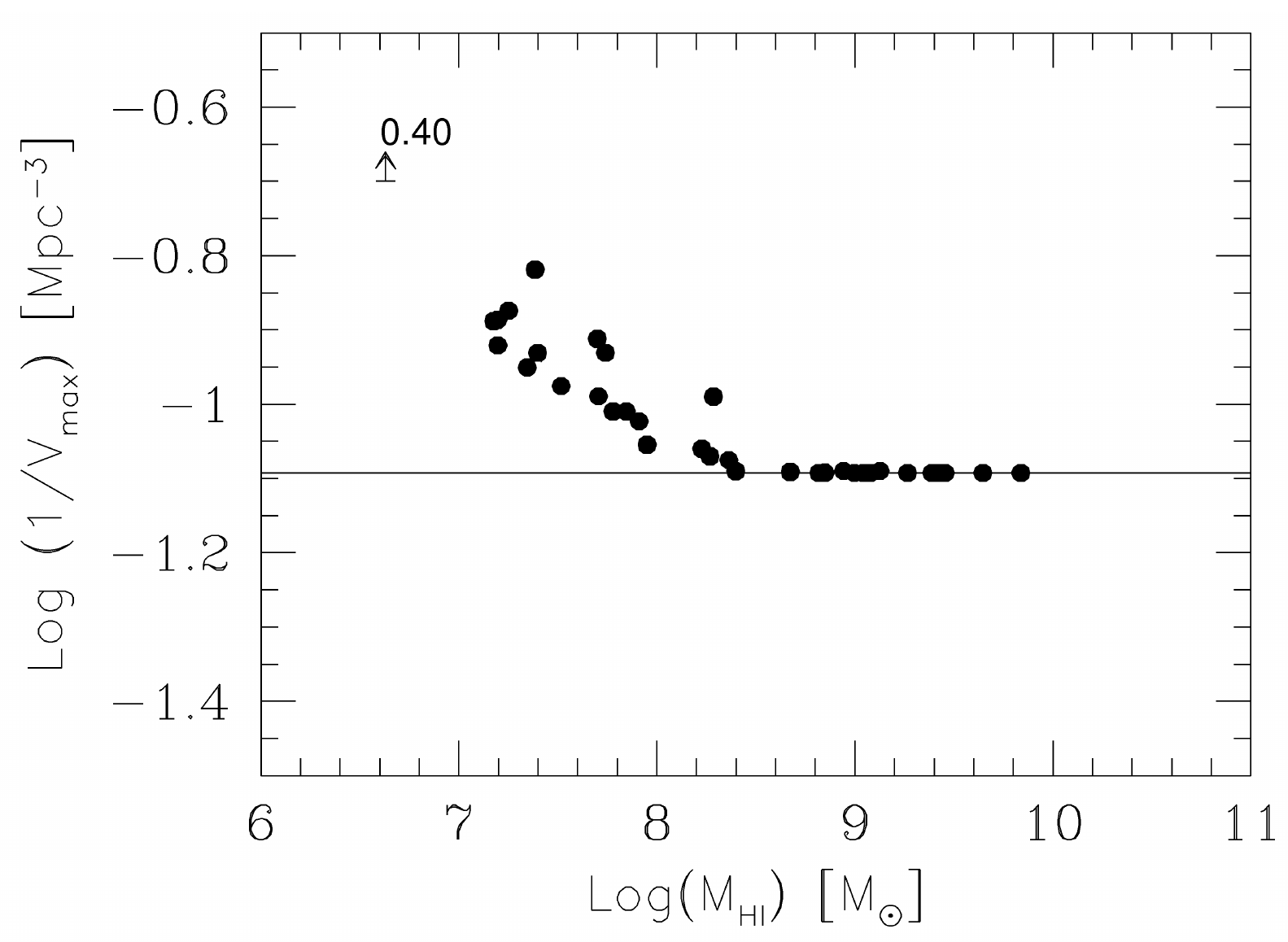}
 \caption{The maximum volume for the galaxies detected in the VLA survey.}\label{fig:volsens}
\end{figure}

For the purpose of calculating $V_{\rm max}$ for each of our HI detections, we note that HI signals are optimally detectable if the resolution of the data matches the spatial extent and the line width of the source. We assume that those dwarf galaxies that can not be detected throughout the entire UMa volume, are spatially unresolved with Gaussian HI line profiles in the smoothed, best-matching data cube in which they are detected at the highest signal-to-noise level as listed in Table \ref{tab:newHIdet}. Furthermore, in order to convert the integrated flux into a total HI mass, we assume that any HI detection is at a distance of 17.4 Mpc. Subsequently, for each HI detection we calculate $D_{\rm max}({\vec x})$ based on the noise in its best-matching data cube, and derive the maximum volume $V_{\rm max}$ within which this galaxy could have been detected at the 6$\sigma$ level, assuming a uniform distribution within the survey volume. For each HI detection in our blind VLA survey, the calculated value of $1/V_{\rm max}$ is plotted in figure \ref{fig:volsens}.

For HI masses larger than $10^{8.4}$ \msol, our survey is clearly volume limited, i.e. the survey volume does not increase for even larger HI masses. Smaller HI masses cannot be detected throughout the entire survey volume, with the result that $V_{\rm max}$ decreases and Log($1/V_{\rm max}$) increases. For the lowest-mass point at $10^{6.6}$ \msol, the correction becomes extremely large. We therefore discard this point from the subsequent derivation of the volume corrected UMa HIMF. It is intriguing, though, that no other reliable HI detection is found in the range Log(\mhi)=6.6$-$7.0 \msol.

\section{The HI Mass Function}

In this section we explain how we derive the HIMF for Ursa Major by combining the targeted WSRT data with the results from our blind VLA-D survey. We also investigate the impact of some of the assumptions we have made and perform a sanity check on our detection rate.

\subsection{Combining the WSRT and VLA-D data}\label{sec:WSRTVLAhimf}

\begin{figure}
\includegraphics[width=0.5\textwidth]{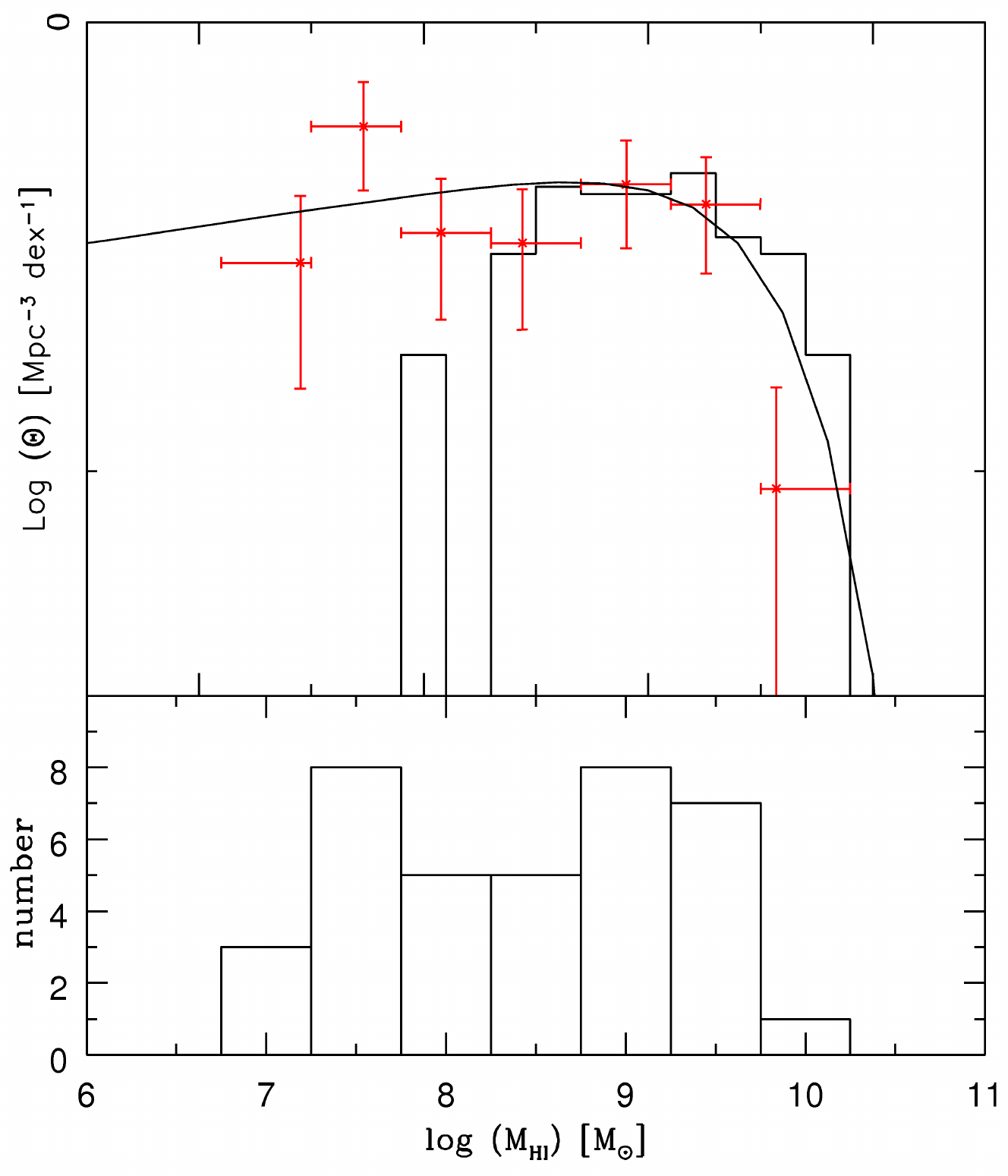}
\caption{The HIMF in Ursa Major as measured with the VLA and the WSRT. {\bf Upper panel}: Horizontal bars show the bin widths, vertical errorbars indicate the Poisson noise. The positions of the points along the horizontal bars indicate the average HI mass of the galaxies within each bin (red dots). The WSRT sample is plotted as a histogram. {\bf Lower panel}: Number of detected galaxies per bin.}\label{fig:HIMFUMa}
\end{figure}

After calculating the volume correction for individual galaxies, it is straightforward to calculate the HIMF. The red data points in the upper panel of Figure \ref{fig:HIMFUMa} show the UMa VLA sample after binning the volume-corrected number of galaxies in 0.5 dex wide mass bins. The vertical error bars are the result of Poisson statistics, whereas the horizontal bars indicate the width of the bins. The horizontal position of a point within a bin indicates the average HI mass of the objects in that bin. The histogram in the upper panel shows the volume-limited survey of Ursa Major carried out with the WSRT by \citet{verheijen2001b} as described earlier. The lower panel shows the distribution of the observed HI masses of the galaxies detected in the VLA survey, assuming all galaxies are at the same distance of 17.4 Mpc.

The volume-corrected HI mass function can be fitted with a \citet{schechter1976} function

 \begin{equation} \label{himf.eq}
 \Theta(\mhi) = \frac{dN}{dlog(\mhi)} = \ln 10 \: \theta^* (\frac{\mhi}{\mhis})^{\alpha+1} \exp-(\frac{\mhi}{\mhis}).
 \end{equation}

The best fit value of the slope $\alpha$, as well as the characteristic HI mass \mhis\ and the characteristic volume density $\theta^*$ that define the knee in the Schechter function, are found by minimizing $\chi^2$ for the expected number of detections in each bin. For the fitting we used the Kapteyn package \citep{terlouw2015} and took an iterative approach in which \mhis\ is based on the WSRT sample and the slope $\alpha$ is derived from the VLA sample. 

The high-mass end of the HIMF of the VLA sample is not very well defined. The surveyed volume is too small to contain a significant number of high-mass HI galaxies. In order to constrain the high-mass end, and thus to constrain \mhis\ , we use the WSRT sample of Ursa Major. We assume this sample is complete in HI above a mass limit of \mhi\ $> 10^{9.0}$ \msol. This HI-complete sample is binned in bins of 0.25 dex to obtain a good constraint on \mhis. The HIMF of this sample is fitted with a Schechter function with a fixed slope of $\alpha = -1.00$. The \mhis\ from this fit is used as a fixed value of \mhis\ in the fit of the VLA sample. The slope $\alpha$ from this fit, is put back as the fixed slope into a new fit to the WRST sample. The value of $\theta^*$ is a free parameter in each fit to allow for a consistent normalisation between the two samples. This procedure is iterated until the fitted values do not change anymore. 

We tested the effect of different binning of the VLA data on the fitted parameters of the HIMF. We chose a bin width of 0.5 dex as a compromise between the Poisson errors for each bin and the number of bins available for fitting. However, we shifted the bins by 0.05 dex to obtain ten different binnings of the VLA data and performed the iterative fitting of the HIMF as described above while keeping the binning of the WSRT data unchanged. Figure \ref{fig:HIMFUMa} shows the Schechter function fit with the first bin centred on Log(\mhi /\msol )=7. The ten fits produced parameters values with ranges of 0.22$<\theta^*<$0.40, 9.8$<$Log(\mhis ) $<$9.8 and $-1.09<\alpha <-0.92$ with weighted mean values of $\theta^* = 0.33 \pm 0.18 (\pm 0.06)$ Mpc$^{-3} dex^{-1}$, Log(\mhis /\msol )=9.8 $\pm 0.83$ and $\alpha = -0.96 \pm 0.13 (\pm 0.06)$. The quoted errors are the quadrature sums of the weighted fitting errors due to the Poisson errors on the data points, and the estimated errors based on the parameter ranges assuming they represent $\pm 1.5\sigma$. The latter are shown in between brackets. We note that \mhis\ was determined by the unchanging WSRT histogram and that the weighted errors are dominated by the Poisson errors on the data points and not by our choice of the bins. The value of $\theta^*$ is based on normalisation against the WSRT sample. 

\subsection{The impact of our assumptions}\label{sec:assumpthimf}

\begin{figure}
\includegraphics[width=0.5\textwidth]{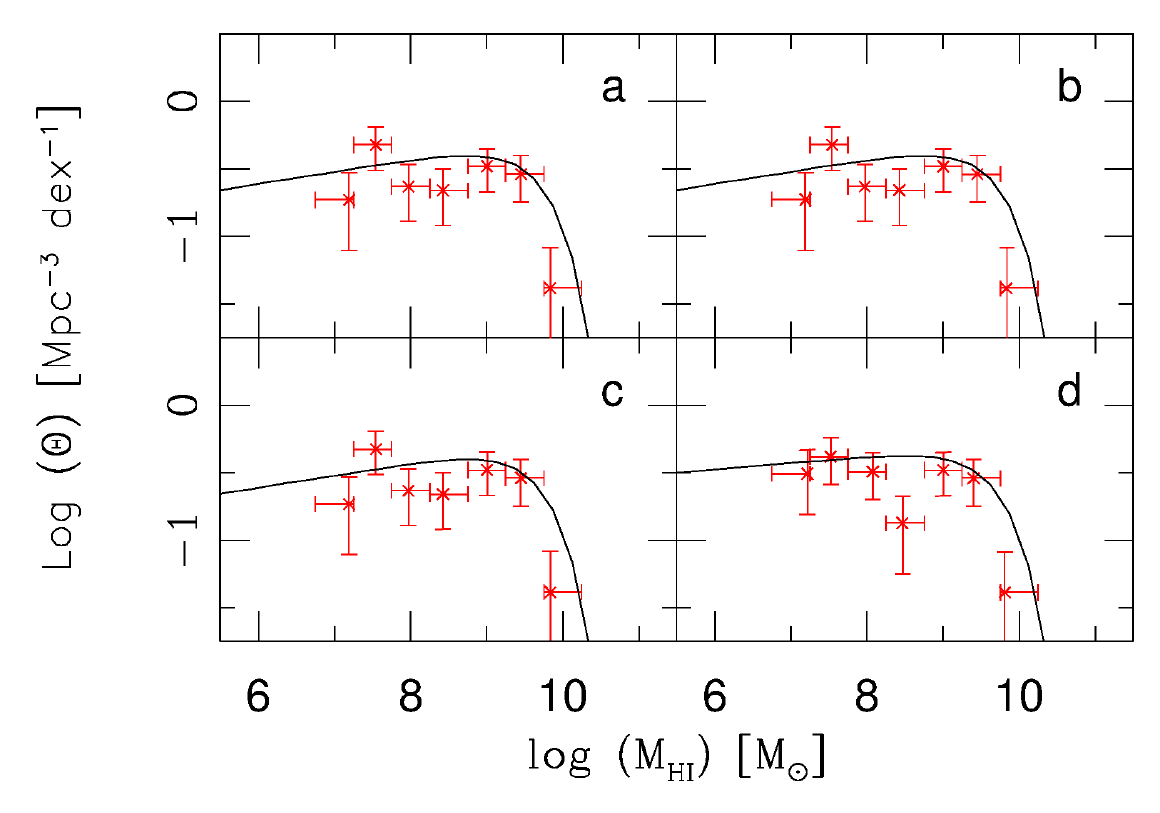}
\caption{The HIMF calculated for different scenarios from Section 2.2: (a) HIMF calculated with the geometrical distribution; (b) HIMF calculated with Tully-Fisher distribution; (c) HIMF calculated assuming all galaxies are approximately at the same distance; (d) HIMF calculated using the Hubble flow to calculate the masses.}\label{fig:diffhimf2}
\end{figure}

The value of $V_{\rm max}$ for an HI detected galaxy depends on its total HI mass and in section \label{sec:volcor} we assumed that all HI detected galaxies are at the same distance of 17.4 Mpc when converting their integrated HI flux into a total HI mass. The assumption that all galaxies are exactly equidistant is, however, an approximation. In fact, we do not know the exact distances of individual galaxies within the UMa volume but we can explore the impact of different assumptions regarding their distances on the slope of the HIMF when calculating their HI masses and corresponding $V_{\rm max}$ values.

\begin{figure}
  \includegraphics[width=0.5\textwidth]{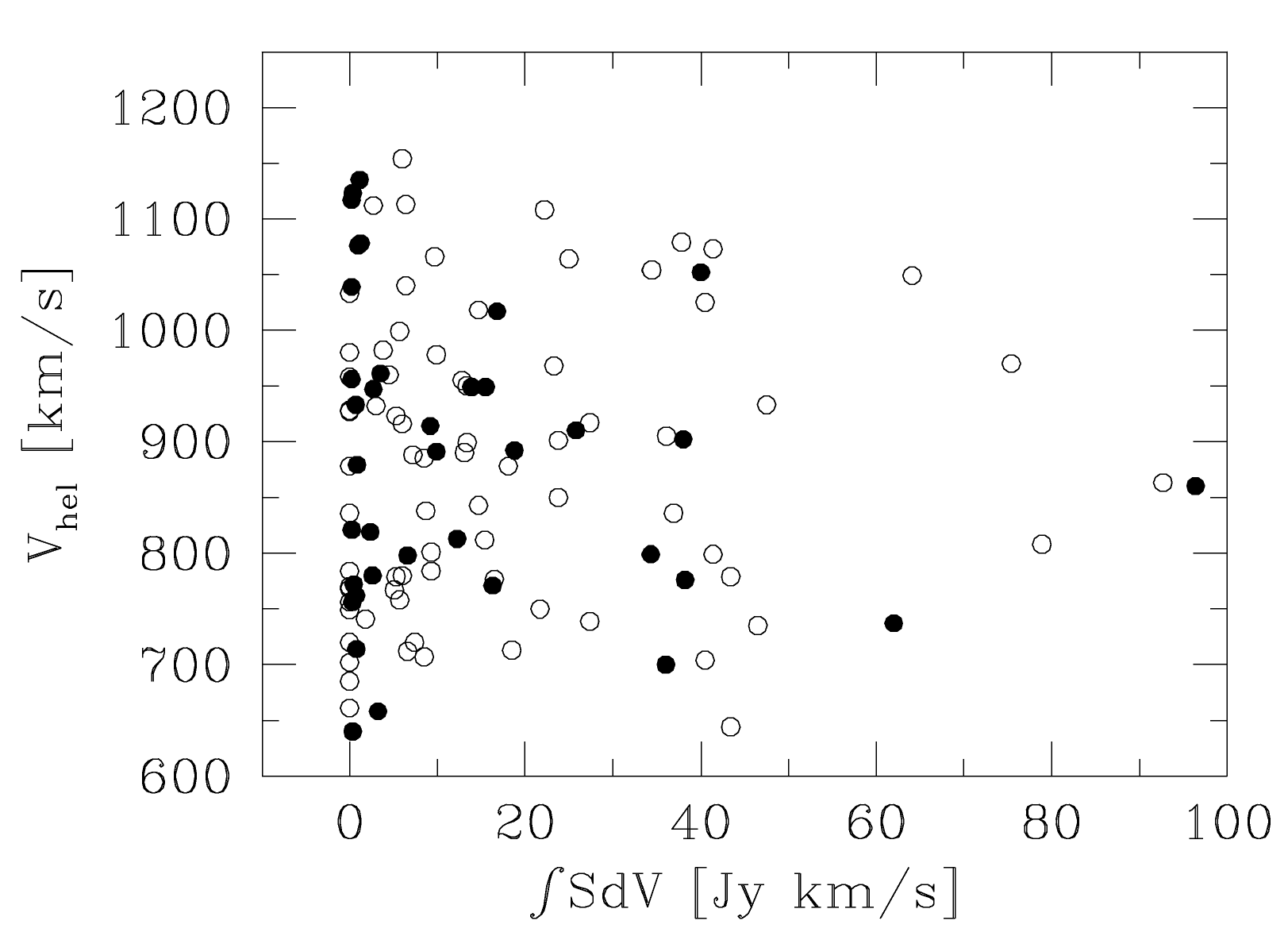}
  \caption{Total flux ($\int SdV$) versus the heliocentric velocity for the galaxies in the VLA sample (black circles) and the WSRT sample (open circles).}\label{fig:W20vsVhel}
 \end{figure}

The UMa volume extends over significant depth, so perhaps the average ensemble distance of 17.4 Mpc to calculate the HI mass of individual galaxies is not appropriate. In large surveys, distance uncertainties are a significant part of the uncertainty of the HIMF. The depth of the UMa region is actually large enough, such that the effects of a Hubble flow could be present within the volume. In Figure \ref{fig:W20vsVhel} $\int SdV$ is plotted against $V_{\rm hel}$ from both the VLA (black circles) and the WSRT sample (open circles), a clear signature of a Hubble flow would be a paucity of objects with small $\int SdV$ and large $V_{\rm hel}$, since dwarf galaxies are expected to be detected only at the near side of the volume. However, in this Figure objects with small $\int SdV$ fluxes appear at all velocities, arguing against the presence of a Hubble flow. Another approach to check the presence of a Hubble flow is by studying the residuals in the TF relation within the UMa region as investigated by \citet{verheijen2001a}. 

Panels 1a and 1b of Figure 8A of \citet{verheijen2001a} illustrate the deviation from the TF relation as a function of the systemic velocity. From this figure a possible correlation is shown between radial velocity and the distance of a galaxy, hinting at the presence of a Hubble flow in the UMa volume. One could wonder what the effect would be if we give galaxies a distance according to their velocity. This HIMF is shown in Figure \ref{fig:diffhimf2} d, $\thetas$ is normalised to represent the number volume density of the entire UMa volume. Its slope is not significantly different from the slope of the HIMF determined using the mean distance to the volume, so a possible effect of the Hubble flow does not change our results.

When using the $1/V_{\rm max}$ volume correction, the distribution of galaxies is assumed to be uniform over the entire Ursa Major volume. Suppose the surveyed volume has a different depth as described in section 2.2, 
we may expect to find a slightly different slope for the HIMF when we use these different depths to calculate the HIMF (adapting $\thetas$ such that it represents the entire volume). The various panels in Figure \ref{fig:diffhimf2}, however, demonstrate that the slope of the HIMF will not be significantly different.

\subsection{A sanity check: comparison with the ALFALFA HIMF}\label{sec:univhimf}

To emphasise that we find a significantly different slope of the HIMF compared to ALFALFA, $-0.96 \pm 0.13$ for UMa versus $-1.37\pm 0.02$ for ALFALFA, we made a more direct bin-by-bin comparison of our detected number of galaxies with the expected number of detected galaxies based on the Alfalfa HIMF, taking into account the survey volume, depth and primary beam attenuation and assuming galaxies are uniformly distributed in the UMa volume. 

The slope of the HIMF of the UMa region is significantly different from the HIMF slopes of HIPASS and ALFALFA. To emphasize this finding, we determine the expected number of detections in case the ALFALFA HIMF were valid for the UMa volume. For this purpose a Schechter function with \mhis\ and slope from the ALFALFA HIMF is used. In order to perform a proper normalisation (\thetas) for this function, we determined \thetas\ by fitting a Schechter function at the highest mass points (around \mhis) of the VLA sample. The noise is not uniform over the whole volume due to primary beam attenuation. To take this into account, we divided the volume into five parts with decreasing gain, so with increasing noise. We determined the fraction of the whole volume that each part occupies. For each part there is a different completeness limit. The completeness limit can be calculated by estimating the integrated flux of a galaxy. Assuming a Gaussian emission line profile, a (faint) galaxy would be detected, if it has an integrated flux density of six times the rms in four channels and two beams. If the gain is $1.0$, this completeness limit would be $S_{int} \sim$ ~ 0.19 Jy km s$^{-1}$. For each part we set an average gain and calculate the corresponding completeness limit. We calculated the number of galaxies in bins of 0.01 dex for each part and place them into this part of the volume under the assumption that galaxies are uniformly distributed over this part of the volume. For each galaxy the flux which we would have observed if the noise was uniform over this part of the volume is calculated. All galaxies with fluxes below the corresponding completeness limit are eliminated. The remaining fluxes are converted back into HI masses in the same way as was done for the VLA sample: using the average distance to the volume (17.4 Mpc).  After having done this for all five parts, we combine the results to obtain the result for the whole volume.

\begin{figure}
\includegraphics[width=0.5\textwidth]{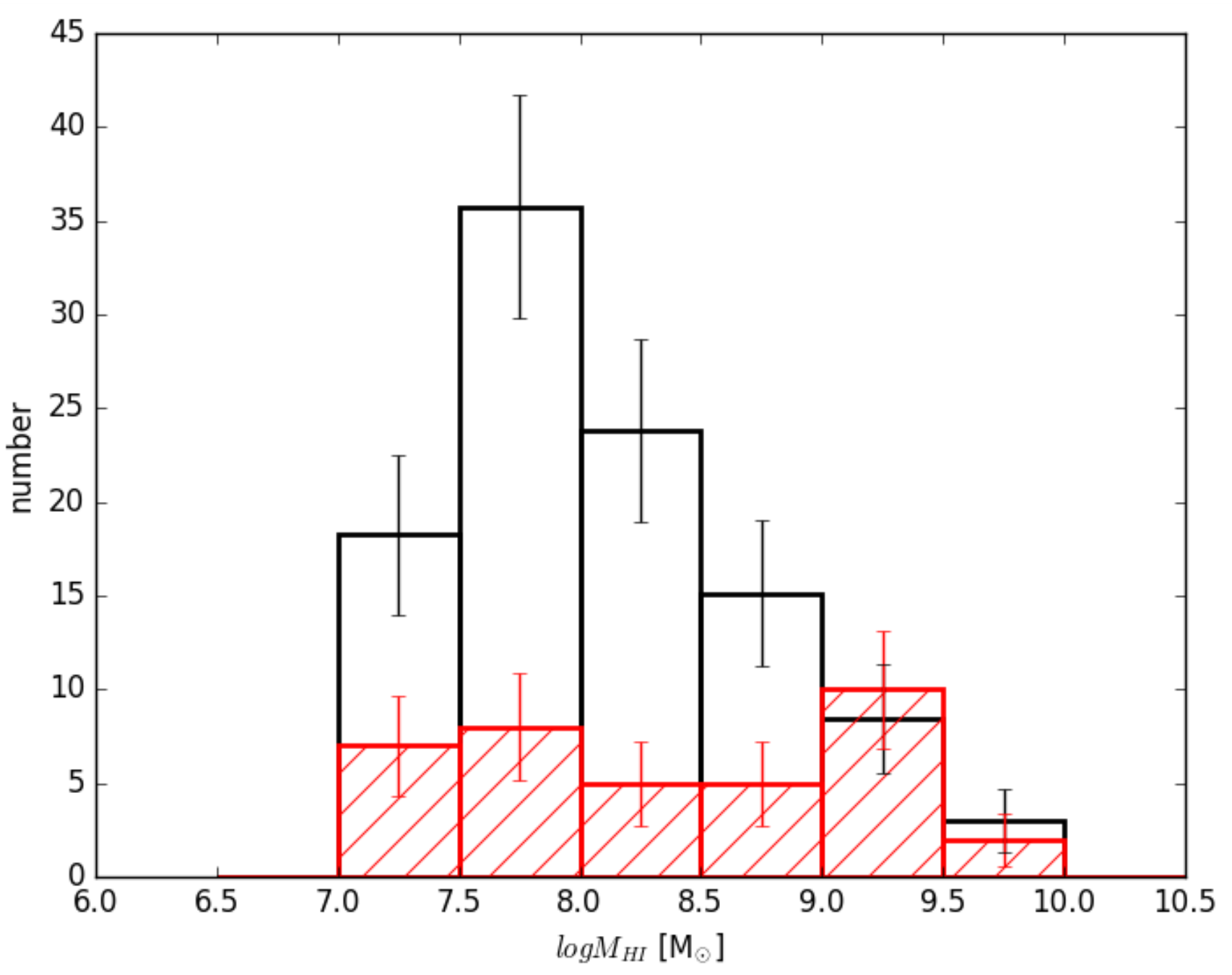}
\caption{The results of distributing the mock galaxies uniformly in the UMa volume, assuming all galaxies are at the same distance, and applying a completion limit (black histogram) and the histogram of the VLA sample (red histogram).}\label{fig:expnum}
\end{figure}

The next step is to bin this data in bins of 0.5 dex to enable comparison with the histogram derived from the VLA observations. The final histogram is shown in Figure \ref{fig:expnum}. This Figure shows the expected histogram based on the ALFALFA HIMF (black) and the histogram from the detected galaxies in the Ursa Major region (red).

Comparison of the histograms reveals a significant difference. At large masses the numbers of galaxies are similar, which is what we would expect, since the slope has no effect on these numbers. We already see a significant difference for the mass bin $10^{8.5}$-$10^{9.0}$, and this difference increases towards lower mass bins. At the lower mass bins the expected number of galaxies if the ALFALFA HIMF were valid within the UMa region is systematically higher than the observed number of galaxies from the VLA sample.

\section{Discussion}

In this paper we determined the HIMF of the Ursa Major volume, with best fit parameters of $\theta^* = 0.19 \pm 0.11$ Mpc$^{-3}$, log $\mhis\ /M_{\odot} = 9.8 \pm 0.8$, and $\alpha = -0.92 \pm 0.16$. 

As shown in Section \ref{sec:univhimf}, we would expect many more detections within the UMa volume if the ALFALFA HIMF were valid. There is, however, another effect besides the volume correction, which could reduce the number of galaxies we would be able to detect and which we did not consider in our analysis. Galaxies are harder to detect in HI when they are edge-on: the linewidth broadens with increased inclination, requiring more flux (i.e. higher HI mass) to rise above the detection threshold. The UMa volume is relatively nearby and the galaxies that could most likely be missed are dwarf galaxies with Gaussian shaped line profiles for which this effect is less severe. We therefore ignored the effect of the inclination of a low mass galaxy on the completeness limit.

\begin{figure*}
\subfloat[Bright galaxies from the catalogue from \citet{pak2014}.]{\label{pg}\includegraphics[width=0.5\textwidth]{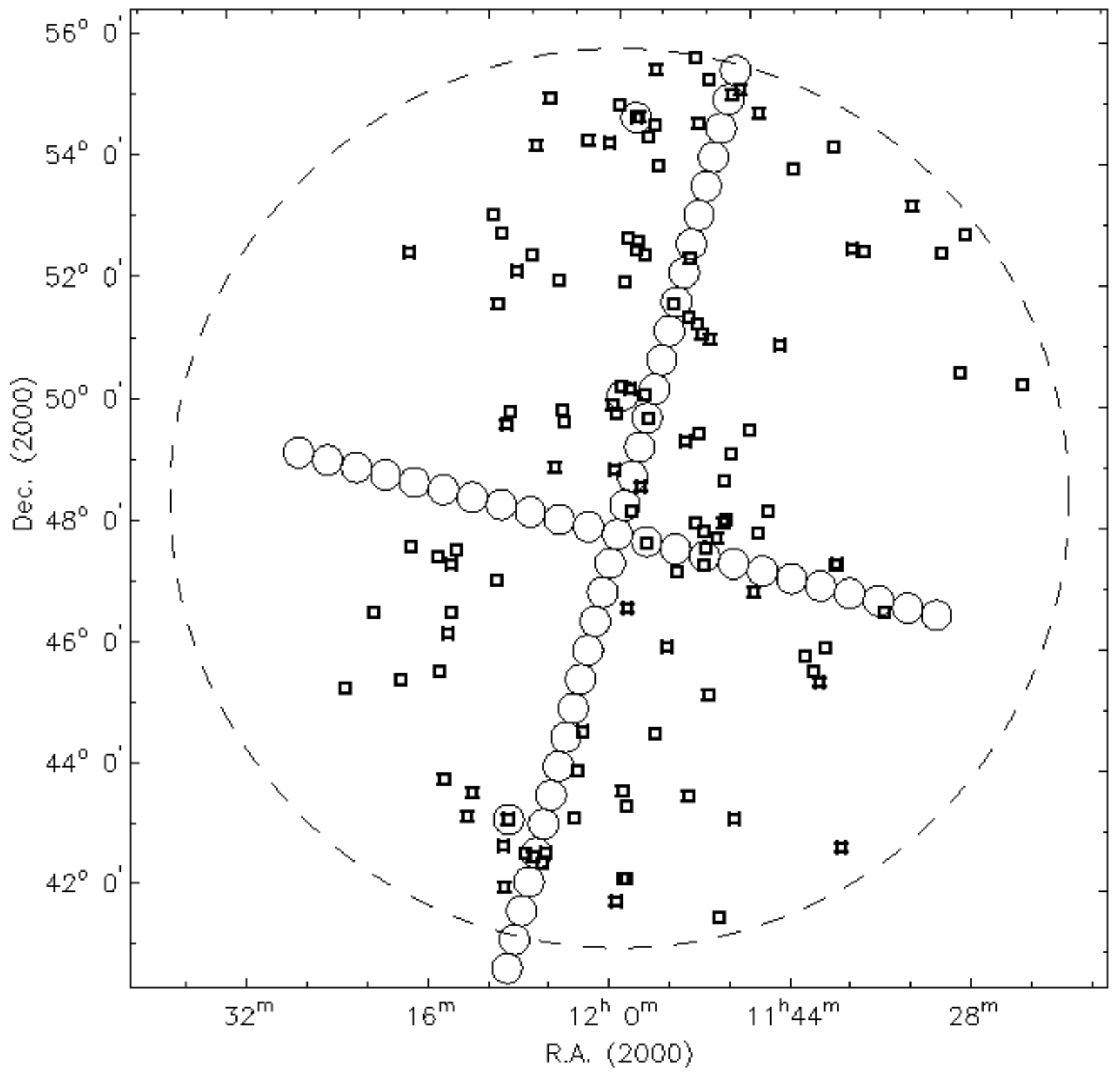}}                
  \subfloat[Dwarf galaxies from the catalogue from \citet{pak2014}.]{\label{pd}\includegraphics[width=0.5\textwidth]{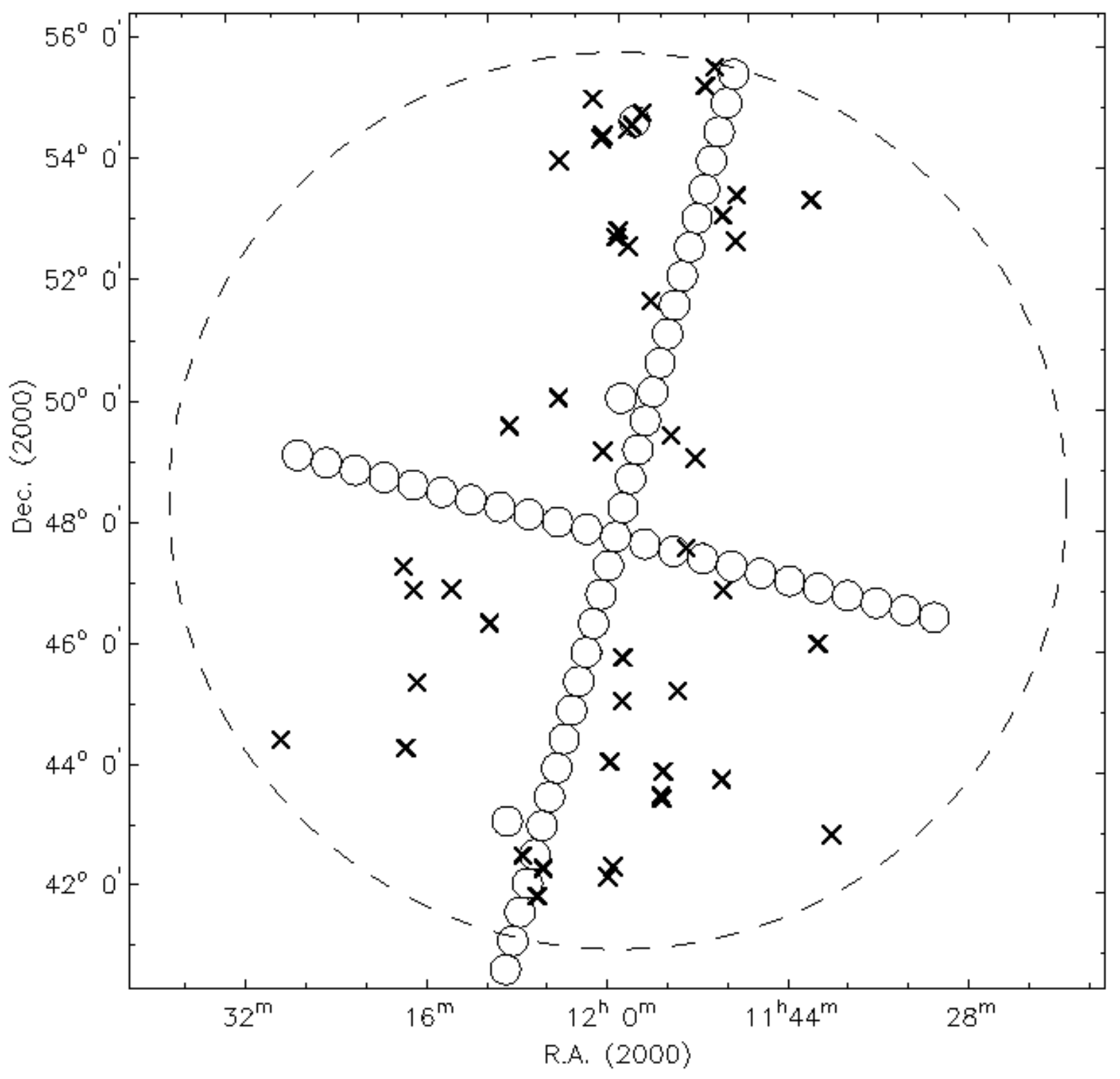}}\\
  \caption{Distributions of the catalogue of the UMa volume by \citet{pak2014}. Galaxies are divided as bright and dwarf galaxies according to the classification from \citet{pak2014}.}\label{fig:pakdistr}
\end{figure*}

Another observational effect is that the galaxies in the observed 16 per cent of the volume may not be representative of the overall galaxy population of the UMa region. \citet{pak2014} compiled a catalogue of the UMa region using redshifts with a completeness of over 94 per cent within a range of $9 < r < 18$ magnitude. We divided the sample from this catalogue into two groups, bright galaxies with M$_{B} < -16.5$ and dwarf galaxies with M$_{B} > -16.5$ and plotted those galaxies (bright galaxies as black squares and the dwarf galaxies as black crosses) in Figure \ref{fig:pakdistr}. It is clear from Figure \ref{fig:pakdistr} that the distribution of the dwarf galaxies follows the distribution of brighter galaxies. Thus the mass function of galaxies in the observed 16 per cent of the volume should be representative for the galaxy population of the entire UMa volume.

Could a non uniform distribution of galaxies in the survey volume, invalidating the $1/V_{\rm max}$ volume correction, explain the discrepancy between the number of expected and observed HI detections as illustrated in Figure \ref{fig:expnum}? Given that the detection limit is lower for low HI mass galaxies in the nearby part of the volume than in the distant part, a higher density of galaxies at a larger distance in the survey volume would qualitatively result in a relatively lower number of low mass galaxies. The detection limit, however, degrades only with a factor two from the near end to the far end of the survey volume while the difference shown in Figure \ref{fig:expnum} is a factor seven. Furthermore, Figure \ref{fig:W20vsVhel} does not hint at a skewed distribution of the recession velocities or a statistically significant differentiation of recession velocities between galaxies of low and high HI mass. Hence, we conclude that a sufficiently strong galaxy density gradient along the line-of-sight to invalidate the $1/V_{\rm max}$ volume correction is extremely unlikely.

Although the HIMF of the UMa region differs substantially from the average HIMF of HIPASS and ALFALFA, it is similar to other results in the literature. Comparing the HIMF of group-like environments, shows that the result of the UMa region is very similar to these results (\citet{pisano2011}; \citet{kovac2007}; \citet{freeland2009}; \citet{deblok2002}; \citet{kilborn2009}). 

An inconsistent result is reported by \citet{stierwalt2009} for the Leo I group. The HIMF slope of the Leo I group is steeper than the slopes from the HIMF of HIPASS and ALFALFA, while its LF is similar to the LF of the UMa region \citep{flint2003}. The galaxy population of the Leo I group is, however, quite different from that in the UMa region. The population of the Leo I group consists for 69 per cent of low-mass objects (45 out of 65 galaxies have \mhi\ $< 10^8$), LSB dwarf-rich galaxies, and three bright L$^{*}$, early-type galaxies (\citet{trentham2002}; \citet{stierwalt2009}). The galaxy population of the CVn groups consists approximately of 55 per cent (38 out of 70 galaxies have \mhi\ $< 10^8$) low-mass objects and is dominated by late-type galaxies. The population of the UMa region is different and consists of more field-like galaxies, with relatively more late-type L$^{*}$ galaxies. Thus, though at first sight the galaxy densities and environments are similar, the Leo I group, the CVn groups and the UMa region appear to have different galaxy populations in terms of gas content and galaxy types and masses. This indicates that besides general environment other properties are instrumental in determining the observed differences. One of these may be the precise (and hard to determine) balance between the different physical processes responsible for galaxy transformation: ram pressure stripping, gravitational interaction and stripping, and star formation induces gas flows.

\cite{zwaan2005} and \cite{springob2005} studied the effect of the environment on the HIMF. They both reported a relation between the galaxy density and the HIMF slope, but interestingly, they found effects in opposite directions. An important and difficult issue here is how to define the environment. Both authors defined the environment by how many neighbours a galaxy has in a catalogue, but based their result on different catalogues. \cite{zwaan2005} use HICAT, an HI catalogue containing all galaxies that have been used to determine the HIMF. \cite{springob2005} used the PCSz (IRAS), which is dominated by dusty and star forming objects. Both catalogues represent different galaxy populations. It is therefore conceivable that they find a different dependence on the environment. Properly qualifying the environment is thus a prerequisite for studying the dependence of the HIMF on environment.

The flat slope of the HIMF could be explained by a lack of low-mass galaxies. Often the explanation for the flattening of the HIMF slope is that this is the result of low-mass galaxies losing their gas in a group-environment (\citet{zwaan2005}; \citet{hess2013}). There are mechanisms that only remove the gas from the galaxies and leave the stellar component undisturbed. If this is the case, we should be able to find those low-mass objects in the optical, hence the LF would be steeper than the HIMF at the low luminosity end. However, the LF of the UMa region derived in \citet{trentham2002} shows that the slope is shallow as well ($\alpha \sim -1.0$). The Local Group shows a similar trend, both the HIMF \citep{pisano2011} and LF \citep{vandenbergh1992} have flat slopes. In the Local Group we also see mechanisms such as ram pressure and tidal stripping, and tidal interactions, but on smaller scales (central galaxy - dwarf galaxy interactions) (\citet{mateo1998}; \citet{spekkens2014}). The flattening of both the LF and the HIMF could be explained by various scenarios: (i) the dark matter haloes never trapped any cold HI gas \citep{klypin1999}, (ii) the galaxies have lost their gas before they start making stars, or (iii) the galaxies have been stripped from their gas 1-2 Gyrs ago, have become `red and dead` \citep{schawinski2014} and are too faint to be catalogued optically. 

\begin{figure*}
\includegraphics[width=\textwidth]{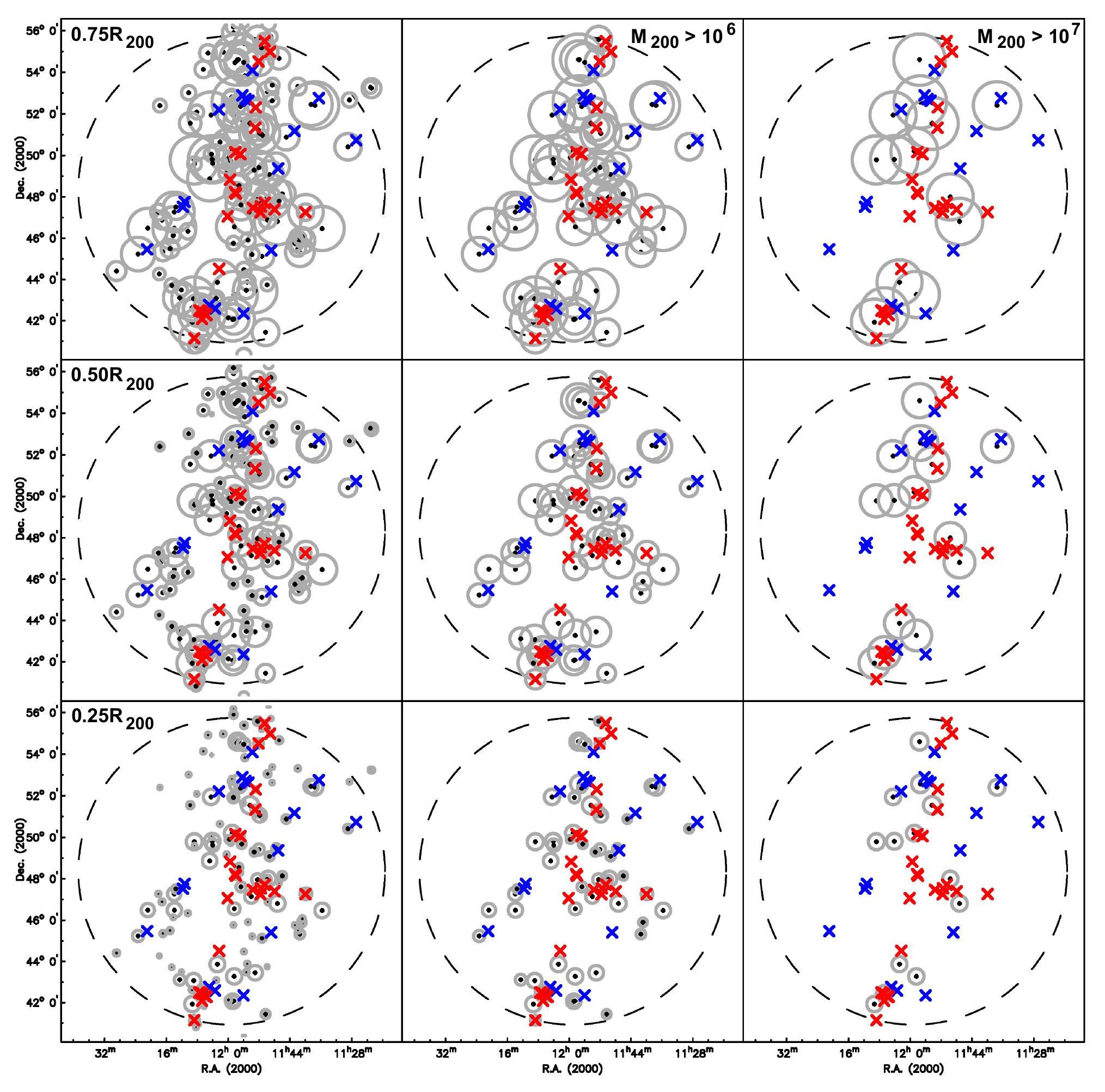}
\caption{The UMa region with for different mass ranges, visualisations of the different size virial zones. The virial zones plotted are from top to bottom: 0.75$R_{200}$, 0.5$R_{200}$, and 0.25$R_{200}$. The mass selection of the plotted galaxies varies from left to right as: all galaxies with their virial zones, only virial zones of galaxies with $M_{200} > 10^{6}$\msol), and only virial zones of galaxies with $M_{200} > 10^{7}$\msol).}\label{fig:virzon}
\end{figure*}

The LF itself also has an environmental dependence. For example \cite{croton2005} showed that the LF becomes steeper in denser environments. Multiple mechanisms that affect both the gas and the stars may be happening in a group-like environment, like stated earlier in the introduction. In the UMa environment, the velocity dispersion is low, which makes ram pressure stripping unlikely, but makes tidal interactions more effective. More detailed HI imaging is needed to determine what mechanism(s) are responsible for the alteration of the galaxy population in groups.

Instead of studying the influence of the cluster environment, group environment, etc. on the steepness of the slope, we can also consider the effect of the sphere of influence (SoI) of individual galaxies, in particular the massive ones, on the slope of the HIMF which probes the smaller objects. A relevant parameter then probably is the fraction of the survey volume that is occupied by SoI’s of galaxies above a certain mass. We can define the SoI of a galaxy as the sphere around a galaxy in which another galaxy can be stripped from its HI gas. This can happen through either ram pressure stripping or tidal stripping, depending on the masses of the galaxies, the velocities of the galaxies with respect to each other, and the presence of X-ray gas in the halo. If a large fraction of the surveyed volume is occupied by SoIs, one would expect a flat to declining HIMF slope, since most dwarf galaxies will be stripped from their HI gas either through ram pressure or tidal stripping. If the dominant stripping process is ram pressure stripping, only the gas would be stripped, not the stars, and the LF would remain steep. If the dominant stripping process is tidal stripping, then the stars could be affected as well, also flattening the slope of the LF. If a small fraction of the surveyed volume is occupied by SoIs, one would expect a steeper slope for both the LF and the HIMF, as dwarf galaxies are not (yet) stripped from their HI gas. 

It would be interesting to test this idea on UMa. Determination of the SoI is difficult, but a crude approximation could be a fraction of $R_{200}$, as it depends on the mass of a galaxy. However, the precise fraction of $R_{200}$ remains unclear, as well as above which mass galaxies actually have a significant SoI. To explore this idea further we plot in Figure \ref{fig:virzon} the UMa region with SoIs assigned to galaxies of different mass ranges (from left to right) and this SoI as a different fraction of $R_{200}$ (from top to bottom). We use the K-band luminosity and the TF relation to calculate for each galaxy a V$_{\text{flat}}$ and assuming that V$_{\text{flat}}$ is similar to V$_{200}$, we then can calculate $R_{200}$ and $M_{200}$. The crosses are all galaxies with  \mhi\ $< 10^9$ \msol\ (which are the galaxies responsible for the low-mass slope of the HIMF), the red crosses are the galaxies present in both the VLA and the WSRT survey, the blue crosses are galaxies present only in the WSRT survey. Figure \ref{fig:virzon} gives a first impression of what fraction of the galaxies with \mhi\ $< 10^9$ \msol\ falls within SoIs for which choice of SoI and $M_{200}$.

\citet{tully2015a} (as well as \citet{tully2015b}) introduced the second turnover radius, $R_{2t}$, the radius of a group within which one would expect the satellite galaxies to be depleted in HI. $R_{2t}$ is approximately half of the $R_{200}$ of the brightest galaxy within a group, so the middle panels from Figure \ref{fig:virzon} roughly indicate the $R_{2t}$'s of the UMa system. If this is the most representative situation, then the majority of the small galaxies are within $R_{2t}$ of at least all galaxies with $M_{200} > 10^{6}$, qualitatively explaining the flat slope of the low mass end of the HIMF. This is what we found in this study. Since both the HIMF and the LF have a flat faint end slope, we could argue that the dominant stripping process in the UMa volume is tidal stripping. With this idea in mind we can also revisit the results for the HIMF in other environments, e.g. the Leo I and CVn I groups. Here the fraction of the volume taken up by significant SoI’s is expected to be much smaller and would lead to a steeper slope of the low mass end of the HIMF as observed in these groups (\citet{stierwalt2009} and \citet{kovac2007}).

\section{Conclusions}
We derived the HIMF of the UMa volume, with best-fitting parameters of $\theta^* = 0.19 \pm 0.11$ Mpc$^{-3}$, log $\mhis\ /M_{\odot} = 9.8 \pm 0.8$, and $\alpha = -0.92 \pm 0.16$. This HIMF slope is consistent with a flat slope ($\alpha = -1.00$), in contrast to the slopes of HIPASS and ALFALFA. It is similar to the slope of the Luminosity Function of the UMa region.

Our data support an environmental dependence of the slope of the HIMF. The UMa volume is an association of many groups with relatively fewer low-mass galaxies. Therefore we can conclude that the galaxy population of groups seems to differ from the average galaxy population. This indicates that the environment may play a role in the evolution of galaxies. In a group environment different processes are dominant compared to the main processes in clusters. Whether this is in the form of feedback or other mechanisms is not clear. 

Another way to quantify environmental effects is to examine the Sphere of Influence (SoI) of individual galaxies, a SoI being a sphere around a galaxy in which another galaxy can be stripped from its HI gas. The slopes of the faint end of the LF and HIMF could depend on the fraction of the observed volume that is occupied by SoIs and on what type of stripping (ram pressure or tidal) is dominant within those SoIs. If ram pressure stripping is dominant, then only the gas is affected (LF remains steep, HIMF flattens), if tidal stripping is dominant, both the gas and the stars are affected (LF and HIMF will both flatten). In the UMa volume both the LF slope and the HIMF slope are flat, suggesting that tidal stripping may be the dominant stripping process within the UMa volume.

\section*{Acknowlegdements}
E. Busekool and J.M van der Hulst acknowledge support from the European Research Council under the European Union's Seventh Framework Programme (FP/2007-2013) / ERC Grant Agreement nr. 291531 (HIStoryNU). M. Verheijen acknowledges support by the Netherlands Foundation for Scientific Research (NWO) through VICI grant 016.130.338.  The National Radio Astronomy Observatory is a facility of the National Science Foundation operated under cooperative agreement by Associated Universities, Inc.

\section*{Data availability}
The data underlying this article will be shared on reasonable request to the corresponding author.

\begin{appendix}
\section{Galaxies newly detected in HI}

\begin{landscape}
 \begin{figure}
  \subfloat{   \includegraphics[width=11cm]{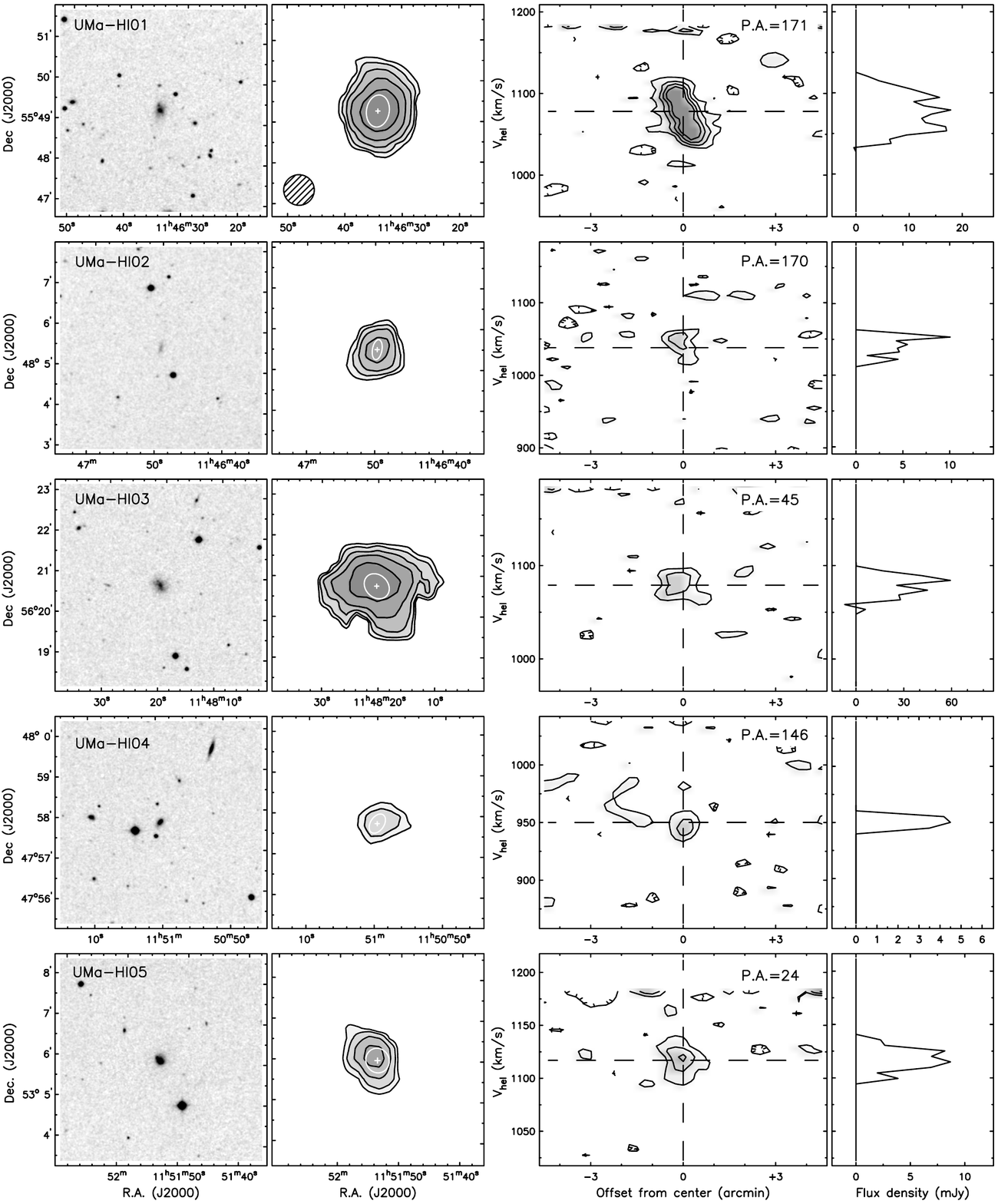}}
  \hfill
  \subfloat{   \includegraphics[width=11cm]{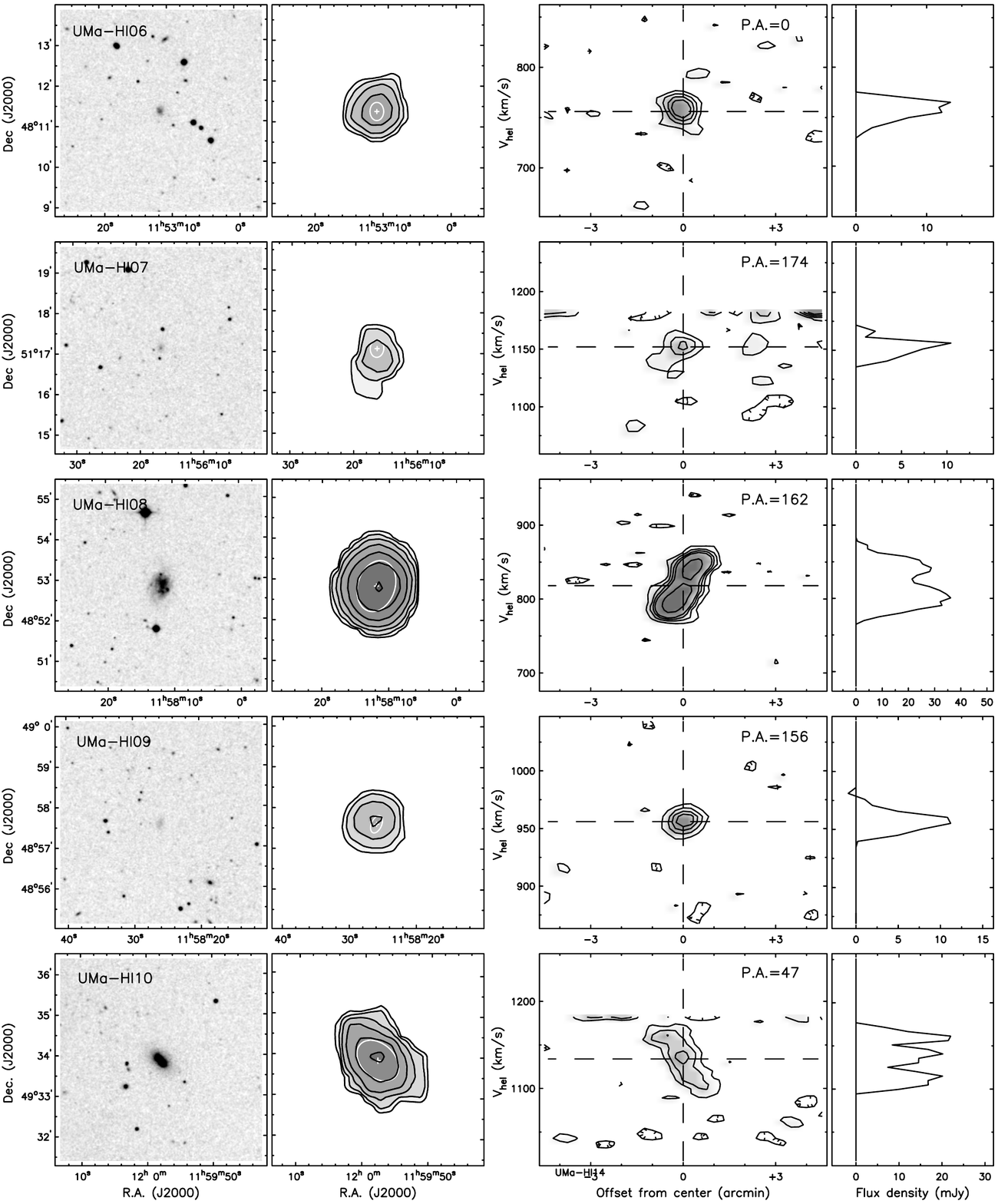}}
 \caption{Galaxies newly detected in HI. Each row shows the data for a single galaxy with from left to right an optical image (DSS), an HI column density map corrected for the primary beam attenuation (contour levels are 1, 2, 4, 8, 16, 32, 64, 128, 256 and 512 $\times 10^{19}$ cm$^{-2}$; the white contour outlines the extent of the optical galaxy), a position-velocity plot along the major axis (contour levels are are -2, 2, 4, 6, 8, 16, 32, 64 and 128 times the rms noise (0.754 mJy/beam for all galaxies except NGC\ 4111 which has an rms noise of 1.43 mJy/beam; the velocity resolution is 10.2 km s$^{-1}$; the position angle is given in the top right corner) and the integrated HI profile.} \label{fig:6a}
 \end{figure}
\end{landscape}

\begin{landscape}
 \begin{figure}
  \subfloat{  \includegraphics[width=11cm]{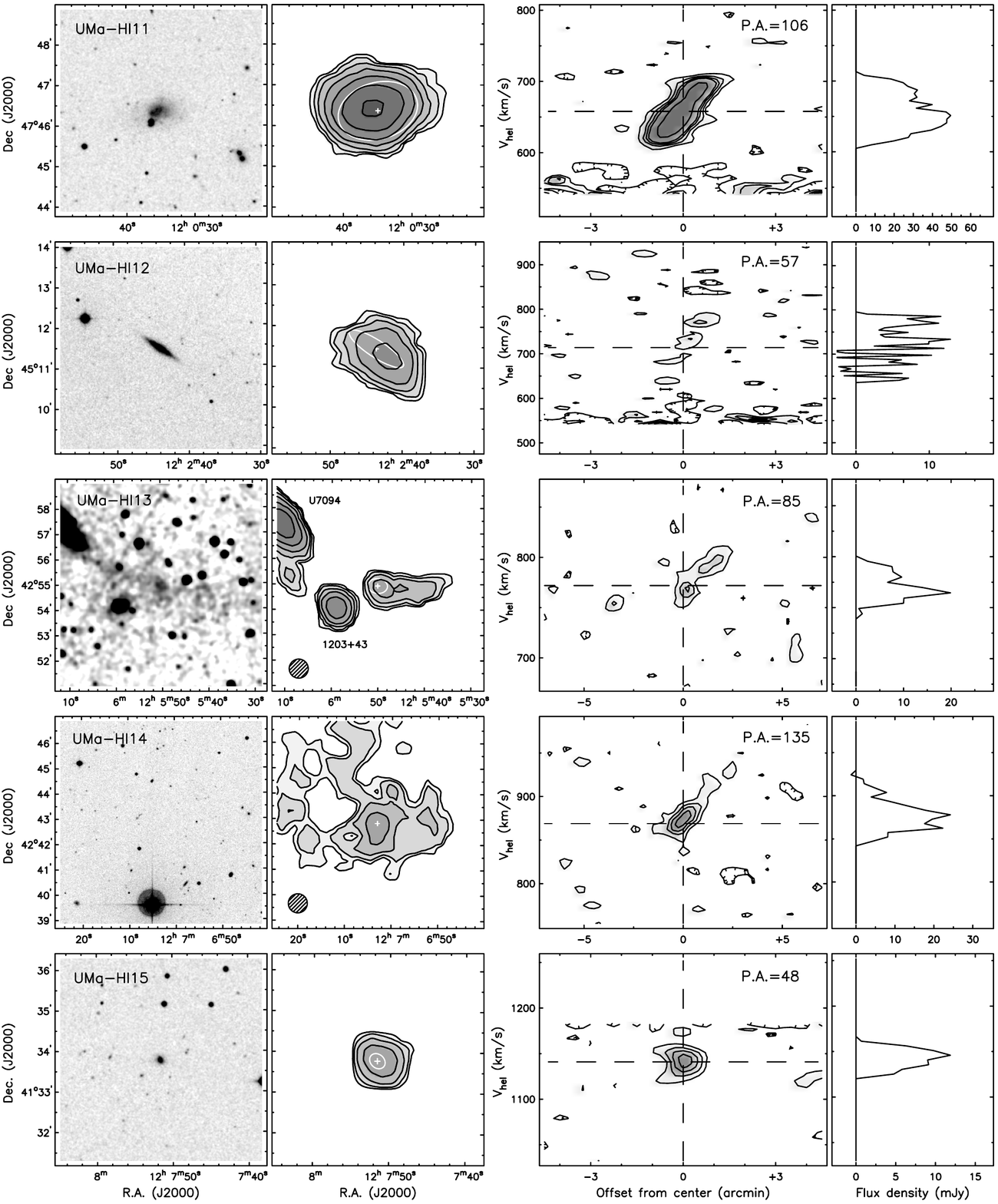}}
  \hfill
  \subfloat{  \includegraphics[width=11cm]{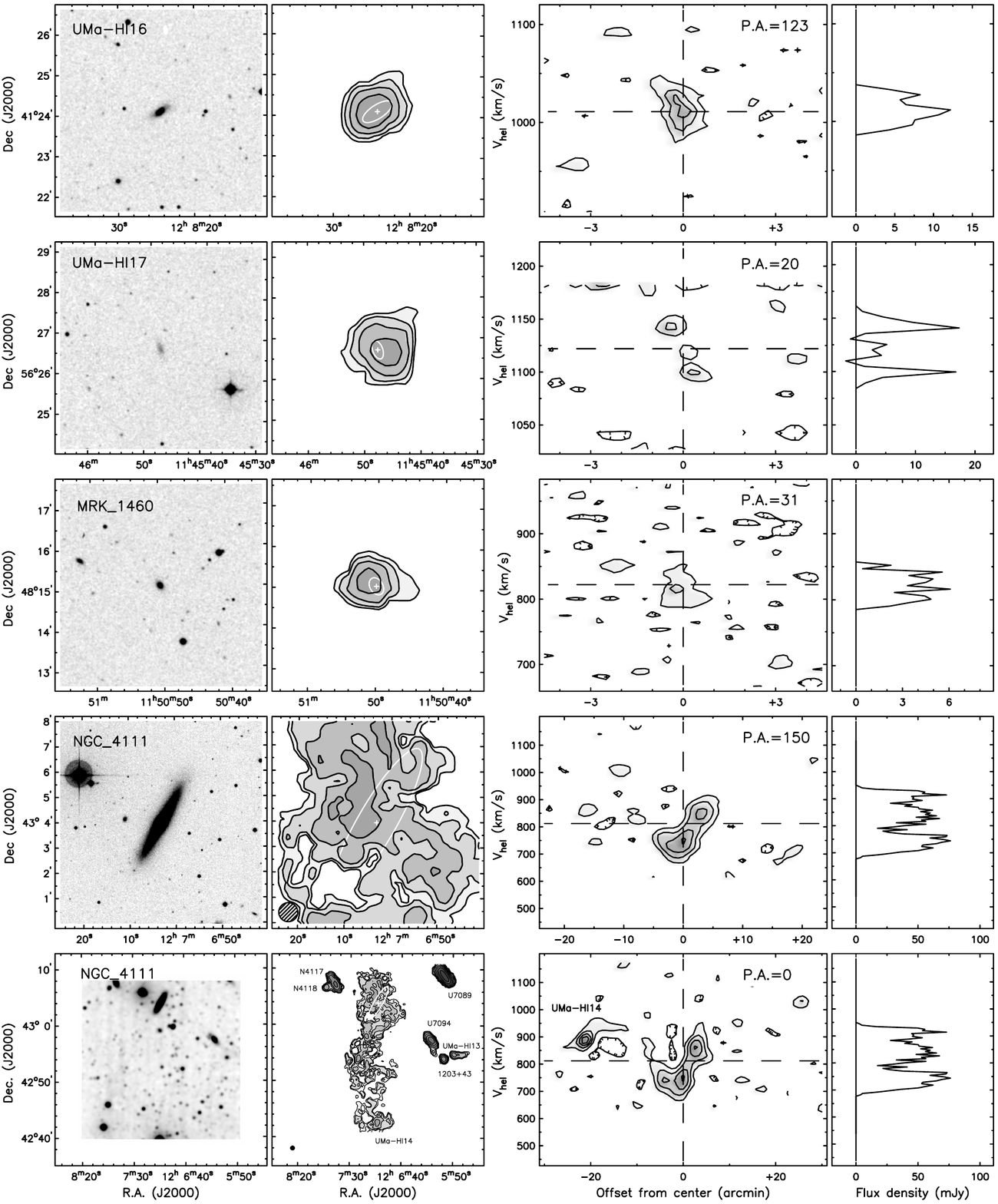}}
 \caption{Galaxies newly detected in HI (continued, for detailed information see figure \ref{fig:6a}).}\label{fig:6b}
 \end{figure}
\end{landscape}

\end{appendix}

\end{document}